\documentclass[10pt,titlepage]{article}
\usepackage[round,numbers,sort,compress]{natbib}

\usepackage[left=1in, right=1in]{geometry}

\usepackage{graphicx}
\usepackage[utf8]{inputenc}
\usepackage{amsmath}
\usepackage{amsfonts}
\usepackage{color}

 \renewcommand{\eqref}{\ref}

\begin{document}
\definecolor{orange}{rgb}{1,0.5,0}
\newcommand{\todo}[1]{{\bf \textcolor{orange}{#1}}}
\newcommand{\change}[1]{{\bf \emph{#1}}}
\newcommand{\obsolate}[1]{{\bf \textcolor{blue}{#1}}}
\newcommand{\ddt}{\frac{d}{dt}}
\newcommand{\playerOne}{\textrm{A}}
\newcommand{\playerTwo}{\textrm{B}}
\newcommand{\dnaP}{\text{DNA}_{\playerOne}}
\newcommand{\dnaG}{\text{DNA}_{\playerTwo}}
\newcommand{\dnaStar}{\text{DNA}_{*}}
\newcommand{\proteinP}{\text{Protein}_{\playerOne}}
\newcommand{\proteinG}{\text{Protein}_{\playerTwo}}
\newcommand{\proteinStar}{\text{Protein}_{*}}
\newcommand{\mRNAP}{\text{mRNA}_{\playerOne}}
\newcommand{\mRNAG}{\text{mRNA}_{\playerTwo}}
\newcommand{\mRNAStar}{\text{mRNA}_{*}}
\newcommand{\nP}{{n_\playerOne}}
\newcommand{\nG}{{n_\playerTwo}}
\newcommand{\nDP}{{d_\playerOne}}
\newcommand{\nDG}{{d_\playerTwo}}
\newcommand{\nMP}{{m_\playerOne}}
\newcommand{\nMG}{{m_\playerTwo}}
\newcommand{\SP}{S_{\playerOne}}
\newcommand{\SG}{S_{\playerTwo}}
\newcommand{\bias}{\text{Bias}_{\playerOne}}
\newcommand{\PP}{P_{\playerOne}}
\newcommand{\PG}{P_{\playerTwo}}
\newcommand{\PZero}{P_{0}}

\newcommand{\discreteP}{{N_\playerOne}}
\newcommand{\discreteG}{{N_\playerTwo}}
\newcommand{\discreteDP}{{D_\playerOne}}
\newcommand{\discreteDG}{{D_\playerTwo}}
\newcommand{\discreteMP}{{M_\playerOne}}
\newcommand{\discreteMG}{{M_\playerTwo}}
\title{Stability and multi-attractor dynamics of a toggle switch based on a
two-stage model of stochastic gene expression}

\author{\textbf{Michael~Strasser}\\
	Institute for Bioinformatics and Systems Biology, \\
	Helmholtz Zentrum M\"unchen\\
	German Research Center for Environmental Health, Germany
	\and \textbf{Fabian~J.~Theis} \\
	Institute for Bioinformatics and Systems Biology,\\
	Helmholtz Zentrum M\"unchen\\
	German Research Center for Environmental Health, Germany\\
	{\small and}\\
	Institute for Mathematical Sciences,\\
        Technische Universit\"at M\"unchen\\
	Garching, Germany
	\and \textbf{Carsten~Marr}\thanks{
           Corresponding author.  Address:
           Institute for Bioinformatics and Systems Biology,
	   Helmholtz Zentrum M\"unchen,
	   Ingolstaedter Landstrasse 1,
	   85764 Neuherberg, Germany,
	   Tel.: +49 (0)89 3187 3642, Fax: +49 (0)89 3187 3585, Email:~carsten.marr@helmholtz-muenchen.de}
	\thanks{Present address: Centre for Systems Biology at Edinburgh, University of Edinburgh, Edinburgh EH9 3JD, UK}
	\\
	Institute for Bioinformatics and Systems Biology, \\
	Helmholtz Zentrum M\"unchen\\
	German Research Center for Environmental Health, Germany}

\date{\today}
\maketitle
\begin{abstract}
A toggle switch consists of two genes that mutually repress each other. 
This regulatory motif is active during cell differentiation and is thought to act as a memory device, 
being able to choose and maintain cell fate decisions. 
Commonly, this switch has been modeled in a deterministic framework where transcription 
and translation are lumped together. 
In this description, bistability occurs for transcription factor cooperativity, while autoactivation leads 
to a tristable system with an additional undecided state.

In this contribution, we study the stability and dynamics of a two-stage gene expression switch within a probabilistic framework inspired by the properties of the Pu/Gata toggle switch in myeloid progenitor cells. 
We focus on low mRNA numbers, high protein abundance and monomeric transcription factor binding. Contrary to the expectation from a deterministic description, this switch shows complex multi-attractor dynamics without autoactivation and cooperativity. Most importantly, the four attractors of the system, which only emerge in a probabilistic two-stage description, can be identified with committed and primed states in cell differentiation.
We first study the dynamics of the system and infer the mechanisms that move the system between attractors using both the quasi-potential and the probability flux of the system.
Second, we show that the residence times of the system in
one of the committed attractors are geometrically distributed.
We derive an analytical expression for the parameter of the
geometric distribution,
therefore completely describing the statistics of the switching process
and elucidate the influence of the system parameters on the residence
time.
Most importantly we find that the mean residence time increases
linearly with the mean protein level. This scaling also holds for a
one-stage scenario and for auto-activation. 
Finally, we study the implications of this distribution for the stability of a switch 
and discuss the influence of the stability on a specific cell differentiation mechanism.
Our model explains lineage priming and proposes the need of either high protein numbers 
or long term modifications such as chromatin remodeling to achieve stable cell fate decisions.
Notably we present a system with high protein abundance that nevertheless requires a probabilistic description 
to exhibit multistability, complex switching dynamics and lineage priming.
\vspace{2cm}

\end{abstract}

\section{Introduction}
\label{sec:intro}
During differentiation, a cell and its progeny cascade through a
number of lineage decisions from stem cells over progenitor cells to
mature functional cells. Many decisions are assumed to be binary and
realized by a toggle switch, a simple cellular memory device.
This network module consists of two genes,
inhibiting each other via mutual promoter binding. In each
differentiating cell, one gene will eventually win this biomolecular
battle, inhibiting the other gene and subsequently activating its
lineage-determining downstream targets.  %
In hematopoiesis, the generation of blood cells, a series of gene
switches has been found to determine the differentiation path of
hematopoietic stem cells and to direct the ratio of mature blood cells
\citep{orkin08_cell_snapshot,Krumsiek2011}. The most prominent
example in this context is the mutual inhibition of Gata-1 and Pu.1,
two transcription factors responsible for the development of erythroid
and myeloid blood cells from common myeloid progenitors
\citep{zhang99_proc-natl-acad-sci-u, arinobu07_cell-stem-cell,
  burda10_leukemia}.

Due to its importance in development, toggle switches are subject to
both experimental and theoretical investigations (for a review, see \cite{macarthur09_nat-rev-mol-cell-bio}).
Using a
deterministic framework under the assumption of large molecule
numbers, \citet{cherry00_j-theor-biol} discussed criteria for working
switches.
More specifically \citet{roeder06_j-theor-biol,  huang07_dev-biol}
and \citet{chickarmane09_plos-comput-biol} used a simple deterministic model
of the toggle switch based on ordinary differential equations in order to
describe the Pu.1--Gata-1 switch in hematopoiesis.
A comprehensive overview and comparison of the different deterministic toggle switch models is provided by \citet{Duff2011}.

All these studies focus on the steady states of the switch and the
parameter dependent bifurcations in a deterministic framework. 
However, protein variations of a
differentiating cell influence the dynamics of the decision making
process and lead to stochastic transitions between the two steady
states. This randomness is induced by gene expression noise, which
has been shown to be ubiquitous in biological systems due to low molecule numbers
\citep{eldar10_nature}.  Thus, the
probabilistic frameworks developed in order to account for gene expression
noise 
(see \citep{paulsson05_physics-of-life-revi} for a review)
have to be applied in order to understand fundamental aspects of toggle switch
properties.

Probabilistic models of the toggle switch account for
low copy numbers and intrinsic fluctuations. In
\citep{kepler01_biophys-j}, the dynamics of an exclusive switch, where
two genes share the same promoter, is discussed within a probabilistic
framework. A comparison of simple switch circuitries is given in
\citet{warren04_physical-review-lett}. Contrary to deterministic
models, transitions between the two macroscopic regimes where one of
the two genes dominates are possible due to the inherently noisy gene
transcription \cite{Schultz2008,Walczak2005a}, even without cooperative binding of transcription
factors \citep{lipshtat06_phys-rev-lett}. More recent contributions
focused on analytic descriptions
\citep{walczak05_biophysical-journal, schultz07_j-chem-phys}, the
switching time between macroscopic regimes for different regulatory
realizations \citep{loinger07_phys-rev-e-stat-nonl,
  barzel08_phys-rev-e-stat-nonl,Schultz2008} or parameter regimes \cite{Walczak2005a},  boundaries for the
switching time \citep{bialek01_}, or delay effects
\citep{zhu07_j-theor-biol}.
Notably, all of these
approaches are based on a one-stage model of gene expression, where
DNA is directly processed into functional proteins. However, it has
been shown that the characteristics of protein noise strongly depend
on the underlying expression model
\citep{thattai01_proc-natl-acad-sci-u,
  shahrezaei08_proc-natl-acad-sci-u}.

In this contribution, we abstract the regulatory details of the prominent myeloid Pu.1-Gata-1 mutual inhibition.
Contrary to common belief, which advocates the lumping of the two stages of expression, 
we show that the inclusion of both mRNA and protein leads to an interesting change in system dynamics.
The probabilistic two-stage description exhibits complex multi-attractor dynamics without autoactivation and cooperativity. 
Remarkably, a recent study reported low numbers of mRNAs in single murine blood cells: 
\citet{warren06_proc-natl-acad-sci-u} found around 10 transcripts of the Pu.1 gene per cell in common myeloid progenitors.
Based on these findings we study a probabilistic description of a toggle switch with low mRNA numbers, 
high protein abundance and in accordance with the known role of Pu.1, monomeric transcription factor binding. 
We deliberately choose the simplest toggle switch model and neglect autoactivation due to our ignorance of the 
logic of activation and inhibition at the promoter. However our results can easily be extended and are discussed 
for the case of dimeric regulation and exclusive autoactivation.

\section{Results}
\label{sec:results}

\subsection{A toggle switch based on a two-stage model of gene expression}
\label{sec:reactions}

\begin{figure}[h]
  \centering
  \includegraphics[width=0.8\textwidth]{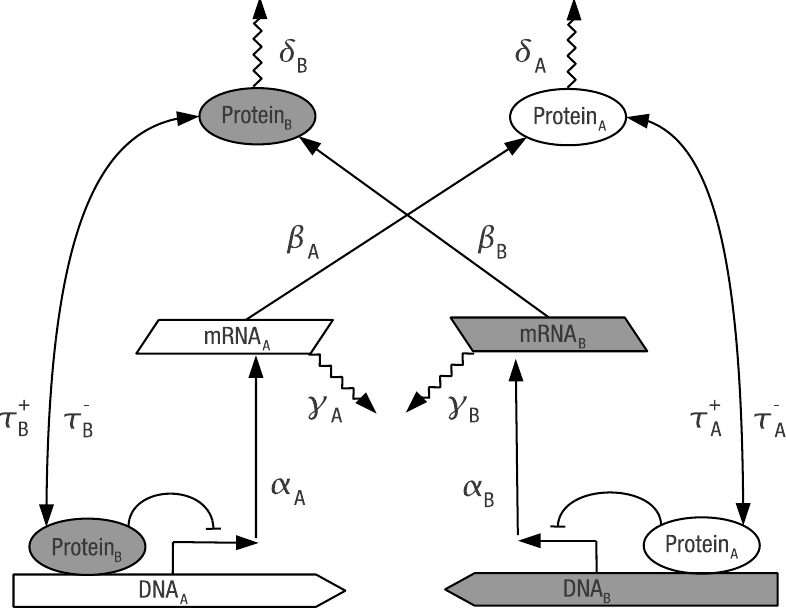}
  \caption{Scheme of the two-stage switch. Species associated with
    gene $\playerOne$ are shown in white, species associated with
    $\playerTwo$ are shown in gray. Solid arrows indicate synthesis
    and binding, jagged arrows indicate degradation. $\mRNAP$ is
    transcribed from $\dnaP$ with rate $\alpha_\playerOne$. It decays
    with rate $\gamma_\playerOne$ and is translated into $\proteinP$
    with rate $\beta_\playerOne$.  $\proteinP$ decays with rate
    $\delta_\playerOne$ and can bind (unbind) $\dnaG$ with rate
    $\tau^+_\playerOne$ ($\tau^-_\playerOne$). Protein-bound DNA leads
    to transcriptional arrest. The topology is symmetric with respect to
    the genes A and B, thus, the same reactions exist for B.}
  \label{fig:model}
\end{figure}

We describe the mutual inhibition of two genes, further on called A
and B, using a two-stage model of gene expression
\citep{thattai01_proc-natl-acad-sci-u,
  shahrezaei08_proc-natl-acad-sci-u} with mutual inhibition being realized
as DNA-protein binding (see Fig.~\ref{fig:model}).
This kind of switch has been implemented in vivo by \citet{gardner00_nature}.
The model can be represented as a set
of biochemical reactions for A and B, respectively, and a set of reaction
rates $\alpha$, $\beta$, etc.:
\begin{tabular}{p{0.45\textwidth} p{0.45\textwidth}}
{\begin{align*}
    &\dnaP \xrightarrow {\alpha_\playerOne} \; \dnaP + \mRNAP\\
    &\mRNAP \xrightarrow{\gamma_\playerOne}\; \emptyset\\
    &\mRNAP \xrightarrow{\beta_\playerOne}\; \mRNAP + \proteinP\\
    &\proteinP  \xrightarrow{\delta_\playerOne}\; \emptyset\\
    &\proteinP + \dnaG \xrightarrow{\tau^+_\playerOne}\; \dnaG^{\text{bound}}\\
    &\dnaG^{\text{bound}} \xrightarrow{\tau^-_\playerOne}\; \proteinP + \dnaG
  \end{align*}}
& \;
{\begin{align}
    \label{eq:13}  &\dnaG \xrightarrow{\alpha_\playerTwo}\; \dnaG + \mRNAG\\
    \label{eq:4} &\mRNAG \xrightarrow{\gamma_\playerTwo}\; \emptyset\\
    \label{eq:5}    &\mRNAG \xrightarrow{\beta_\playerTwo}\; \mRNAG + \proteinG\\
    \label{eq:9}    &\proteinG  \xrightarrow{\delta_\playerTwo}\; \emptyset\\
    \label{eq:10}    &\proteinG + \dnaP \xrightarrow{\tau^+_\playerTwo}\; \dnaP^{\text{bound}}\\
    \label{eq:11} &\dnaP^{\text{bound}}
    \xrightarrow{\tau^-_\playerTwo}\; \proteinG + \dnaP 
  \end{align}}
\end{tabular}
\\\noindent
Reactions \ref{eq:13} and \ref{eq:4}
correspond to mRNA
transcription from an unbound promoter and mRNA degradation,
respectively. Reactions \ref{eq:5} and \ref{eq:9} resemble
protein translation and degradation.  The last two reactions
\ref{eq:10} and \ref{eq:11} describe the binding and unbinding of
a protein to the antagonistic gene and thereby the transition from an
active to an inactive promoter and vice versa.  Bound DNA lacks
the ability to be transcribed.
We would like to emphasize that here $\tau^+$ and $\tau^- $ are rates rather than times.
Note that we assume monomeric transcription factor binding as the simplest of regulatory interaction 
(which has recently been shown to be able to induce bimodal gene expression \cite{Lipshtat2006}).
Our system's topology is
symmetric with regard to the two genes, and so are the two columns of
reactions \ref{eq:13}--\ref{eq:11}
upon the exchange of gene labels A
and B.

This model of gene expression is a highly simplified abstraction of the complex processes in the cell. 
Condensing transcription into a single biochemical reaction does not account for the various steps required to transcribe a gene, 
e.g. the assembly of the transcription initiation complex, 
unwinding of DNA or transition of the polymerase to elongation phase. 
Postprocessing and transport mechanisms are also neglected. 
However, simplified models of gene expression have successfully been applied to experimental data, 
supporting the validity of these simplifications \cite{Harper2011,Raj2006,huang07_dev-biol}.

Most commonly one will study the properties of the system in a deterministic framework using ordinary differential equations (ODEs) 
that describe the time-evolution of species concentrations \citep{roeder06_j-theor-biol,  huang07_dev-biol,chickarmane09_plos-comput-biol,cherry00_j-theor-biol}.
The ODEs can directly be inferred from 
reactions \ref{eq:13}--\ref{eq:11}
assuming mass action kinetics:

\begin{tabular}{p{0.45\textwidth} p{0.45\textwidth}}
{\begin{align*}
    \ddt \nDP &= \tau^-_\playerTwo  (1-\nDP) - \tau^+_\playerTwo  \nDP  \nG\\
    \ddt \nMP &= \alpha_\playerOne  \nDP - \gamma_\playerOne  \nMP\\
     \ddt \nP &= \beta_\playerOne  \nMP - \delta_\playerOne  \nP \\ 
     &+ \tau^-_\playerOne (1-\nDG) - \tau^+_\playerOne  \nDG  \nP\\
  \end{align*}}
& \;
{\begin{align}
      \ddt \nDG  &= \tau^-_\playerOne  (1-\nDG) - \tau^+_\playerOne  \nDG  \nP\label{eq:ode1}\\
      \ddt \nMG  &= \alpha_\playerTwo  \nDG - \gamma_\playerTwo  \nMG\label{eq:ode2}\\
      \ddt \nG  &= \beta_\playerTwo  \nMG - \delta_\playerTwo  \nG  \label{eq:ode3}\\
     &+\tau^-_\playerTwo (1-\nDP) - \tau^+_\playerTwo  \nDP  \nG \notag
  \end{align}}
\end{tabular}

\noindent where $d_*$ is the abundance of unbound $\dnaStar$, 
$m_*$ is the abundance of  $\mRNAStar$ and $n_*$ is the abundance of $\proteinStar$ for $*\in \{A,B\}$.

Bound DNA is expressed in terms of unbound DNA due to mass conservation. %
Solving Eq.~\eqref{eq:ode1}, \eqref{eq:ode2} and \eqref{eq:ode3} at steady state by setting all time derivatives to zero  
yields two solutions, one being biologically irrelevant due to its negative species abundances.
Given non-negative initial conditions the system will always converge towards the positive steady state solution \citep{Mueller-Herold1975}, given by (see Supporting Material for details)
\begin{align}
\nMP^{(ss)} &= \nMG^{(ss)} = -\frac{\delta \tau ^- }{2 \beta  \tau ^+} \left(1  -\eta\right)\label{eq:odeSol1}\\
\nP^{(ss)} &= \nG^{(ss)} = -\frac{\tau ^-}{2 \tau ^+}\left(1 -\eta\right)\label{eq:odeSol2}\\
\nDP^{(ss)} &= \nDG^{(ss)} = \frac{2}{1+\eta}\label{eq:odeSol3}
\end{align}
with $\eta = \sqrt{\frac{4 \alpha  \beta  \tau ^+}{\gamma \delta \tau^-}+1}$.
All parameters are positive and for simplicity assumed to be symmetric 
for players A and B ($\alpha = \alpha_\playerOne = \alpha_\playerTwo, \ldots$).

We now assess the stability of the positive solution Eq. \eqref{eq:odeSol1} - \eqref{eq:odeSol3} using standard linear stability analysis. 
To reduce the complexity of our system for the stability analysis, we apply a quasi steady state approximation to the DNA binding/dissociation process ($\dot{\nDP} = \dot{\nDG} = 0$), 
reducing the dimensionality of our system to four equations.
We evaluate the corresponding Jacobian at the single positive solution
and use the Hurwitz criterion to verify that all its eigenvalues have negative real part. 
We conclude that the system has one stable positive fixed point but we cannot analytically exclude the existence of limit cycles. 
However, inspection of the system's phase portrait (see Fig.\;S4 in the Supporting Material) indicates that no limit cycles exist.
Summarizing, we showed that the deterministic model has only one steady state solution and is thus monostable.

However, since the deterministic approach is only valid in the limit of large numbers, small molecule numbers of DNA, mRNA, and possibly proteins advocate a discrete
probabilistic description of the toggle switch.  We define the state
of the system at time $t$ as a vector $x(t)$, where $x_i(t)\in \mathbb{N}_0$ is the
abundance of species $i$ at time $t$. 
Note that the state space is discrete as opposed to the deterministic model.
To emphasize this difference we use the uppercase notation $\discreteDP$, $\discreteDG$, $\discreteMP$, $\discreteMG$, $\discreteP$, $\discreteG$ 
for the number of molecules of the respective species.
We can describe how the
probability ${\cal P}(x,t)$ of being in a certain state $x$ changes
over time by using the master equation of the system \citep{kampen92_}
\begin{align*}
  \dot{{\cal P}}(x,t) = \sum_{x'} \left[w_{xx'} {\cal P}(x',t) - w_{x'x}
  {\cal P}(x,t)\right]\;.
\end{align*}
The first term considers transitions from states $x'$ with rate
$w_{xx'}$ to state $x$, while the second term accounts for transitions
from $x$ to all other possible states $x'$ with transition rates
$w_{x'x}$. The transition rates, also called propensities \cite{gillespie07_annu-rev-phys-chem} are
determined by the reaction rates and the number of reagents of the
corresponding reactions (see Supporting Material 2 for an
explicit form of the master equation of the system). 

Even though the master equation describes the dynamics of the system more accurately than ODEs 
(most obviously for low particle numbers) it is still an approximation of cellular dynamics as it assumes 
spatial homogeneity inside a cell and does not account for time delays. 
Still, the protein distribution predicted by the master equation of a two-stage expression model was 
indeed observed experimentally \cite{Taniguchi2010a}, supporting the stochastic two-stage model.

Since the master
equation for the switch is analytically solvable only for a number of
approximations (see e.g. \cite{walczak10_arxiv-preprint-arxiv}) and not
integrable for large molecule abundances, we simulate the system
trajectories using Gillespie's algorithm \citep{gillespie1977ess}. Each
trajectory follows the master equation, and the set of infinite
trajectories constitutes the distribution that solves the master
equation.
To obtain appropriate parameters values for stochastic simulation,
we delineate upper bounds for synthesis parameters from biophysical
arguments and adapt degradation parameters to fit desired molecular
levels. Table S1 in the Supporting Materials lists all
used parameter values.
Additionally, the analysis below has been conducted for a second, differently motivated, 
parameter set (see Fig.\;S3 in Supporting Material) and yields qualitatively identical results.
\newcommand{\force}{outflux\;}

\subsection{Dynamics and quasi-potential}
\label{sec:regimes}
\begin{figure}[h]
  \centering
  
    \includegraphics[width=0.8\textwidth]{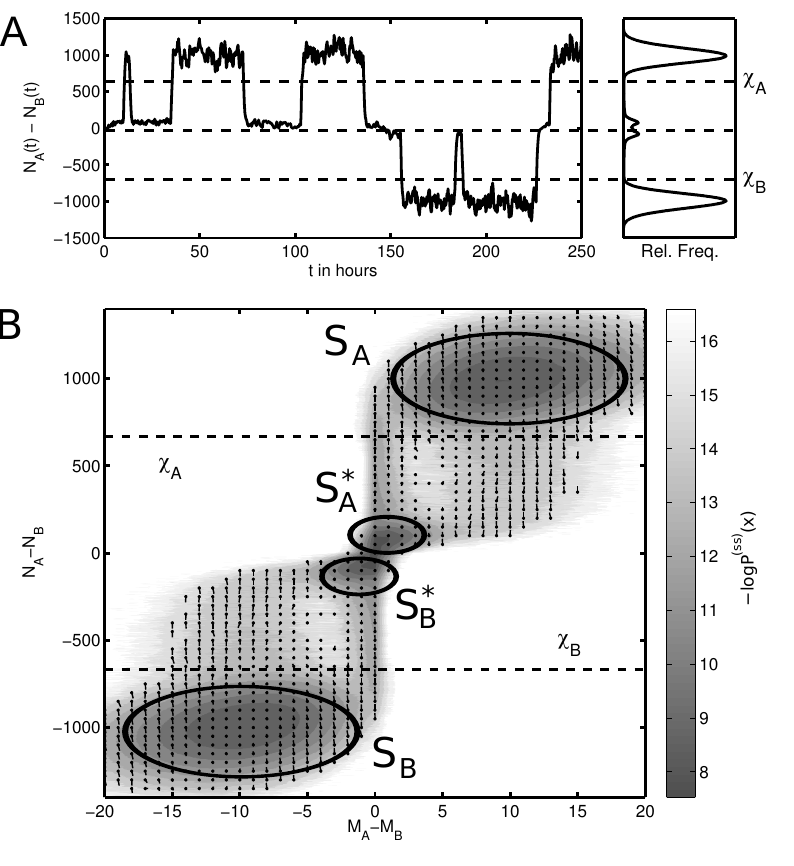}
  \caption{  Dynamics and quasi-potential of the switch showing the different attractors of the system.
  (A) The timecourse of $N_\playerOne(t) - N_\playerTwo(t)$
    clearly shows the dominating attractors, which can be separated in state
    space via the thresholds $\chi_\playerOne$ and $\chi_\playerTwo$.
    Either A dominates (attractor $\textrm{S}_\textrm{A}$), or B dominates (attractor
    $\textrm{S}_\textrm{B}$), or the system is temporarily locked by two bound
    promoters with only marginal protein expression of A or B (attractors $S^*_\playerOne$ and $S^*_\playerTwo$).
    A histogram of $N_\playerOne(t) - N_\playerTwo(t)$ is shown on the right.
    (B) The
    quasi-potential, defined as $U(x) = - \mathrm{log} {\cal P}^{(ss)}(x)$, includes the mRNA dimension of the system. 
    It shows the four
    possible attractors as basins in a probability landscape. $\textrm{S}_\textrm{A}$ and $\textrm{S}_\textrm{B}$ are visible as basins at the lower left and upper right corners, 
    whereas $\textrm{S}^*_\textrm{A}$ and $\textrm{S}^*_\textrm{B}$ are located around the origin ( $N_\playerOne - N_\playerTwo = M_\playerOne - M_\playerTwo =0$) of the landscape.
    Additionally, the \force $F(x)$ acting on the system at the state $x$  in state space are indicated as lines (circles correspond the the origin of the vector). 
    Note that the \force is different from concept of deterministic field lines.
    These vectors show that there are different paths for entering and leaving the dominating attractors.
    Parameters for the simulation are given in Table S1.
}
  \label{fig:regimes}
\end{figure}

In this section we discuss the main features of the switch dynamics. 
Contrary to the deterministic model, time courses of the stochastic toggle switch model show multistable behavior (Fig.~\ref{fig:regimes}A).
Given the parameters in Table S1 our toggle switch can adopt different
attractors: 
The two attractors where one player dominates the other  (called  $S_\playerOne$ and $S_\playerTwo$ depending on which player dominates) are clearly visible in Fig.~\ref{fig:regimes}A.
A careful inspection of the timecourse and the probability distribution in Fig.~\ref{fig:regimes}A shows that there also exist two intermediate attractors 
where protein numbers are similar ($N_\playerOne-N_\playerTwo \approx 0$). These attractors are called $S^*_\playerOne$ and $S^*_\playerTwo$ from now on.
In the timecourses of the system (Fig.~\ref{fig:regimes}A) one observes that the system frequently switches between the dominating and the intermediate attractors. 

To get a deeper understanding of the complex dynamics of the system the notion of a quasi-potential can be used.
The quasi-potential $U$ of the system  is calculated through the relation
$U(x) = - \log {\cal P}^{(ss)}(x)$, where ${\cal P}^{(ss)}(x)$ is the steady state distribution of the system. 
The number of dimensions of the state space where the quasi-potential is defined equals the number of species in the system.
Here the probability ${\cal P}^{(ss)}
(x)$ of a state $x$ in steady state is estimated from 15000 stochastic simulation runs obtained by the Stochkit software toolkit \citep{Li2008}.
In Fig.~\ref{fig:regimes}B the projection of the quasi-potential on the $N_\playerOne-N_\playerTwo$, $M_\playerOne-M_\playerTwo$ plane is shown.
The four attractors $S_\playerOne$, $S_\playerTwo$, $S^*_\playerOne$ and $S^*_\playerTwo$ can be seen clearly in the quasi-potential of the system.
The two attractors $S_\playerOne$ and $S_\playerTwo$ appear as basins at the lower left and upper right corner of Fig.~\ref{fig:regimes}
whereas the intermediate attractors $S^*_\playerOne$ and $S^*_\playerTwo$ are located at the center and are not well separated.
The dominating attractors can easily be distinguished from the intermediate attractors via parameter dependent thresholds $\chi_\playerOne,\chi_\playerTwo$ in the protein dimension 
(see Supporting Material 4).

Importantly, one has to keep in mind that the system considered is out of equilibrium and that the dynamics of a non-equilibrium system  
are not entirely determined by the gradient of the quasi-potential but by an additional 
curl flux stemming from the non-integrability  of the system \citep{wang08_proc-natl-acad-sci-u}.
As a consequence barrier heights in the quasi-potential do not necessarily correlate with the probability of crossing the barrier.

To understand the dynamics of the switch in more detail we therefore consider for each state $x$ in the state space the \force $F(x)$  acting on the the system at this point \cite{Schultz2008}.
We calculate the \force as: 
\[F(x) = {\cal P}^{(ss)}(x) \sum_y {\cal P}(y|x)  (y-x)\;,\] %
where the probability ${\cal P}(y|x)$ of state $y$ succeeding state $x$ and the probability ${\cal P}^{(ss)}(x)$ are calculated from stochastic simulations.
Note that the \force is different from the concept of field lines used in phase portraits of ordinary differential equations.
The \force $F(x)$ is plotted as small arrows in  Fig.~\ref{fig:regimes} (vectors are normalized and circles correspond the origin of the vectors) for all states $x$ with ${\cal P}^{(ss)}(x) > 2.5 \cdot 10^{-7}$. 
This indicates where the system will move from the current state on average.
Due to this \force the system enters and leaves the attractors $S_\playerOne$ and $S_\playerTwo$ through different paths. 
This phenomenon has been described in \cite{wang10_biophys-j} and linked to the emergence of time directionality in non-equilibrium systems.
In order to move from high ($S_\playerOne$ or $S_\playerTwo$) to low ($S_0$) protein numbers, at first the corresponding mRNA number has to drop.
On the contrary, moving from low to high protein numbers requires the rise of mRNA numbers first.

\begin{figure}[h]
  \centering
  \includegraphics[width=0.8\textwidth]{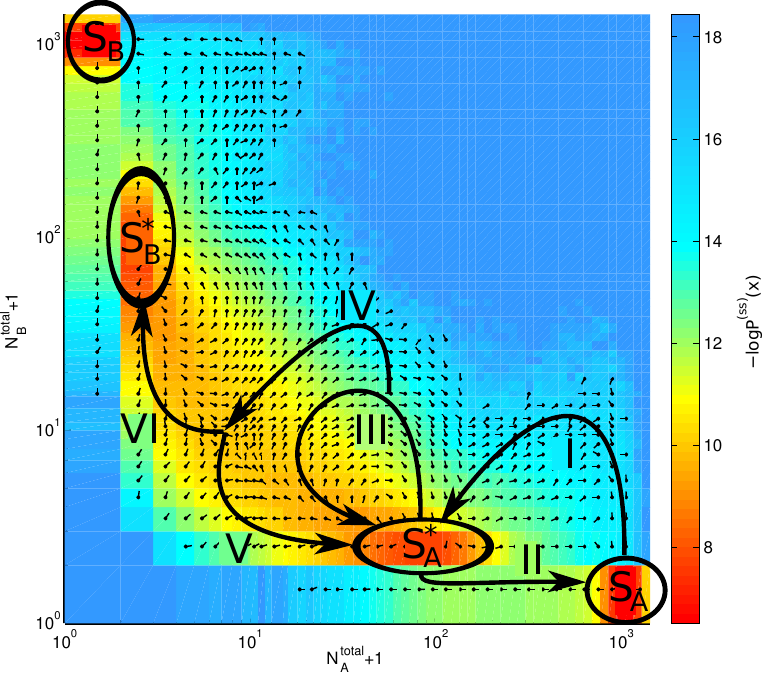}
  \caption{Quasi-potential of the system projected onto the $N_\playerOne^{\textrm{total}}$ and $N_\playerTwo^{\textrm{total}}$ dimensions. 
	  Note that both axis are on logarithmic scale and are shifted by 1 in order to include $N_\playerOne^{\textrm{total}}=0$ and $N_\playerTwo^{\textrm{total}}=0$. 
	  Therefore the lowest row in the plot corresponds to the case $N_\playerTwo^{\textrm{total}}=0$.
	    The quasi-potential $U(x) =  -\log {\cal P}^{(ss)}(x)$ is color coded where red areas reflect minima of the landscape.
	    Visible are four minima corresponding to $S_\playerOne$ (lower right), $S_\playerTwo$ (upper left), $S_\playerOne^*$ (lower middle) and $S_\playerTwo^*$ (middle left).
	  The  vectors of the \force 
	  at each point in state space are drawn as lines (circles correspond the the origin of the vector). Note that the \force is different from concept of deterministic field lines.
	  In contrast to Fig.~\ref{fig:regimes} the vectors are normalized and therefore show only the direction, not the magnitude of the field.
	  Bold arrows reflect typical trajectories (I-VI) of the system. For a discussion, see the main text.
}
  \label{fig:arrowsApp}
\end{figure}

A different view on the system's dynamics is provided by the quasi-potential landscape and \force in the $N_\playerOne^{\textrm{total}},N_\playerTwo^{\textrm{total}}$ plane (Fig.~\ref{fig:arrowsApp}), 
where $N_\playerOne^{\textrm{total}} = (1 - \discreteDG) + \discreteP$ is the total number of $\proteinP$ in the system, bound to DNA (first term) or free (second term).
Choosing $N_\playerOne^{\textrm{total}}$ and $N_\playerTwo^{\textrm{total}}$ as projected dimensions shows four distinct basins in the quasi-potential landscape.
Two basins correspond to the attractors $S_\playerOne$ and $S_\playerTwo$. These are characterized by high amounts of the dominating protein and zero proteins of the repressed species.
The attractors $S^*_\playerOne$ and $S^*_\playerTwo$ are now clearly separated. %
In these two basins a single protein of one species is present and only a moderate protein number of the other species.
In the following we show why these basins emerge and how the system moves between the attractors.

We explain the dynamics of the system with a typical trajectory of the system:
Let us start with the trajectory in the attractor $S_\playerOne$ (lower right) where $\proteinP$ dominates $\proteinG$.
Due to stochastic fluctuations in the promoter status, eventually a burst of proteins of $\playerTwo$
will occur and inhibit the promoter of $\playerOne$, whose protein numbers will drop (Fig.~\ref{fig:arrowsApp}, trajectory I).
While the formerly dominating $\proteinP$'s are degraded, also the newly created $\proteinG$ quickly decreases in 
numbers and only one bound $\proteinG$ is saved from degradation.
This drives the system towards the origin in the quasi-potential of Fig.~\ref{fig:arrowsApp}.
However, a single $\proteinG$ cannot completely suppress the promoter of $\dnaP$, leading to a small but constant synthesis of $\proteinP$.
The system settles into an intermediate state ($S_\playerOne^*$) defined by the presence of one $\proteinG$ 
and an intermediate amount of $\proteinP$ originating from the leaky inhibition of $\dnaP$ and bursting.
In order to leave this basin the system has two options: 
Either the single $\proteinG$ is degraded when it momentarily is not bound to the promoter. 
Consequently the levels of $\proteinP$ rise again and the system reaches $S_\playerOne$. 
The system is moved to the lower border of the quasi-potential where a strong \force pushes it towards $S_\playerOne$ (Fig.~\ref{fig:arrowsApp}, trajectory II). 
Alternatively, a burst of $\proteinG$ displaces the system from $S_\playerOne^*$ into regions 
where the vector field points strongly towards the diagonal $\discreteP^{\textrm{total}} = \discreteG^{\textrm{total}}$ (Fig.~\ref{fig:arrowsApp}, trajectory III).
However this burst is typically not strong enough to move the system onto the diagonal and it will fall back into the basin $S_\playerOne^*$.
In order to enable a change from $S_\playerOne^*$ to $S_\playerTwo^*$ the system has to reach the diagonal. 
This is accomplished if, while the system is moving towards the diagonal after the burst, 
additional bursts of $\proteinG$ move it onto the diagonal (Fig.~\ref{fig:arrowsApp}, trajectory IV).
Once the system has hit the diagonal both protein levels will drop to very low numbers since non of the players has any significant advantage.
Here by chance the system will move to any side of the diagonal and either towards $S_\playerOne^*$ or $S_\playerTwo^*$ (Fig.~\ref{fig:arrowsApp}, trajectories V, VI).

We find that leaving $S_\playerOne^*$ towards $S_\playerOne$ (Fig.~\ref{fig:arrowsApp}, trajectory II) is much more probable than hitting the diagonal from $S_\playerOne^*$ (Fig.~\ref{fig:arrowsApp}, 
trajectory IV), which would provide the chance of switching.
This is obvious from the mechanism described above: 
Even though the events triggering the two alternatives (degradation of $\proteinG$ and an initial burst of $\proteinG$) have similar probabilities, 
the diagonal crossing requires additional events and is therefore much less probable.
This cannot be deduced from the quasi-potential landscape alone:  From Fig.~\ref{fig:arrowsApp} it can visually be inferred that the barrier separating $S_\playerOne$ and $S_\playerOne^*$ is higher than the 
barrier separating $S_\playerOne^*$ and $S_\playerTwo^*$. This wrongly suggests that moving between $S_\playerOne^*$ and $S_\playerTwo^*$ occurs more frequently 
than moving between $S_\playerOne$ and $S_\playerOne^*$.

Comparing the system dynamics of our switch with other descriptions we find that 
(i) deterministic one-stage and two-stage models show no bistability while 
(ii) a probabilistic one-stage model exhibits tristability with only one intermediate attractor
(see Fig.\;S2 in the Supporting Material and \cite{Lipshtat2006}).
We speculate that translational bursting destabilizes the intermediate attractor of the one-stage model, 
where none of the two players can overwhelm the other.
Bursting provides an easy mechanism to escape this deadlock situation: It gives the player whichever bursts first a huge advantage over the other, 
giving rise not only to one protein (as in the one-stage model) but several proteins.
As a result, the two-stage system is always quickly pushed away from the diagonal and stabilizes in the attractors $S_\playerOne^*$ or $S_\playerTwo^*$.
Thus, only the combination of a probabilistic description with a two-stage model of gene expression leads 
to the complex multi-attractor dynamics described above.
\subsection{Residence times}
\label{sec:switching-time}
Genetic toggle switches are thought to be 
involved in the differentiation process of cells. 
A common idea is that different cell fates correspond to the different attractors of the system \citep{Waddington1957}.
Therefore it is of interest how long the system will stay in one of these attractors.
In this contribution, we focus on the time the system will stay in the attractors $S_\playerOne$ or $S_\playerTwo$.
We assume that only in these two attractors the concentration of either player is sufficiently 
high to carry out a downstream biological function which resembles the switch's decision.

In previous contributions, such quantities
have been calculated or determined by stochastic simulation for simpler
switch models and were called spontaneous switching time \cite{bialek01_}, switch
lifetime \citep{warren04_physical-review-lett}, mean
first-passage time \citep{kepler01_biophys-j}, or switching time \citep{barzel08_phys-rev-e-stat-nonl}. 
Since the switch may flip from
a dominating to an intermediate attractor, we choose residence time as the
appropriate term for the quantity calculated below.
In the following, we derive an
analytical approximation for the time the switch stays in a
dominating attractor, $S_\playerOne$ or $S_\playerTwo$, called the
residence time $t_s$. 
A simulation study for $S_\playerOne^*$/$S_\playerTwo^*$ suggests qualitatively similar behavior (see Fig.~S5 in the Supporting Material).

Let us assume that the system is in attractor $\SP$.  Hence, the promoter of
$\dnaG$ is bound by $\proteinP$ while the promoter of $\dnaP$ is
unbound. We assume that the protein levels in this attractor can be
described with the simple two-stage model 
\citep{thattai01_proc-natl-acad-sci-u}, resulting in a mean
$\proteinP$ level of $\bar{N}_\playerOne = (\alpha_\playerOne
\beta_\playerOne) / (\gamma_\playerOne \delta_\playerOne$). Consequently,
the protein level of $\proteinG$ is $N_\playerTwo = 0$ as it is inhibited by the high levels of $\proteinP$.  In order to
leave $\SP$ it is crucial that one $\proteinG$ is synthesized, which
then can bind the promoter of $\dnaP$ and shut down the synthesis of
$\proteinP$, ultimately driving the system out of $S_\playerOne$ and into $\SP^*$.  This
trajectory (called trajectory I in Fig.~\ref{fig:arrowsApp}) involves the following events: (i) unbinding of $\proteinP$
from $\dnaG$, (ii) synthesis of $\proteinG$ during the unbound phase,
and (iii) binding of $\proteinG$ to the promoter of $\dnaP$ before
$\proteinG$ is degraded.

First we describe the unbinding of $\proteinP$ from $\dnaG$. While the
system is in $\SP$, $\proteinP$ dissociates various times, leaving
the promoter of $\dnaG$ unbound.  The average time the promoter remains
unbound, $t_u$, is equal to the average time until a binding reaction
occurs, which is
\begin{align*}
  t_u = \frac{1}{\tau^+_\playerOne \cdot
    \bar{N}_\playerOne}\;.\label{eq:meanField1}
\end{align*}
The time the promoter stays unbound is a random variable itself, but
for simplicity we approximate it with its mean value.  Note that $t_u$
depends, somewhat counterintuitively, on $\tau^+$ and not on $\tau^-$,
with $\tau^+_\playerOne \bar{N}_\playerOne$ being the propensity
for a binding reaction. Again we emphasize that $\tau^+$ and $\tau^-$ are rates (rather than times).

To ultimately synthesize a $\proteinG$, at least one $\mRNAG$ has to
be transcribed during $t_u$ and translated before degradation.  The
probability of $k$ transcription reactions to happen during $t_u$ is
\begin{equation*}
\label{eq:1}
  P_\textrm{Poisson}(K=k)= \frac{(\alpha_\playerTwo \cdot t_u)^k}{k!}\cdot
\exp(-\alpha_\playerTwo \cdot t_u)\;,
\end{equation*}
as the number of transcription reactions $K$ during $t_u$ is
Poisson-distributed with mean $\alpha_\playerTwo \cdot t_u$.  Thus,
the probability of at least one transcription during the unbound phase
is
\begin{align*}
  q_{s} = 1 - P(K=0) = 1 -
  \exp\left(-\frac{\alpha_\playerTwo}{\tau^+_\playerOne \cdot
      \bar{N}_\playerOne}\right)\;.\label{eq:snythesisProb}
\end{align*}
The probability of translation during an average mRNA lifetime $1/
\gamma_\playerTwo$ is accordingly $q_{t} = 1 - \exp\left(- \beta_\playerTwo /
  \gamma_\playerTwo \right)$. Finally the probability for a binding
reaction during average protein lifetime $1/\delta_\playerTwo$ is $q_{b} = 1 -
\exp\left(-\tau^+_\playerTwo /
  \delta_\playerTwo\right).\label{eq:bindProbTwo}$ 

However, not only one but several unbound phases may occur before
$\proteinG$ is successfully synthesized.  The number $L$ of unbound
phases until and including successful synthesis follows a geometric distribution,
$P(L=l) = (1-q)^{l-1} q$ with parameter $ q= q_s \cdot
q_t \cdot q_b$.  The average number of unbound phases during a time
interval $\Delta t$ is $\tau_\playerOne^- \cdot \Delta t$. Thus, we
can convert the random variable $L$ into $T = L /\tau_\playerOne^-$
via a linear transformation of a random variable, giving the actual
time until successful synthesis of $\proteinG$. Notably, the
derivation of the distribution for residence times goes beyond previous
mean-field approximations. Using the properties of the geometric
distribution for the random variable $T$, we end up with the mean and
the variance of the residence time:

\begin{equation}
\label{geometric}
  t_s  =  \frac{1}{\tau^-_\playerOne \cdot q_s q_t q_b} \;\;\;
  \mathrm{and} \;\;\;
  \sigma^2_{t_s} =\frac{1}{(\tau^-_\playerOne)^2}
  \cdot \frac{1-q_s q_t q_b}{(q_s q_t q_b)^2}\;.
\end{equation}

An important approximation for the residence time can be derived under
the assumption of rapid translation and slow mRNA degradation, $\beta
\gg \gamma$, leading to $q_{t} \approx 1$. This implies that it is quite
certain that an mRNA will be translated at least once before
degradation. In the regime of rapid transcription factor binding
($\tau^+ \gg \delta, \alpha $) the probability for a binding reaction
is close to one, $q_b \approx 1$, while the probability for at least
one transcription can be approximated with $q_s \approx
\alpha_\playerTwo / (\tau_\playerOne^+ \bar{N}_\playerOne)$. Taken together,
this leads to a linear dependence of the residence time on the
protein number, 

\begin{equation}
t_s \approx (\tau^+_\playerOne/\tau^-_\playerOne) \cdot
(\bar{N}_\playerOne/\alpha_\playerTwo) \; .\label{eq:3}
\end{equation}

\begin{figure}[h]
  \centering
  \includegraphics[width=0.8\textwidth]{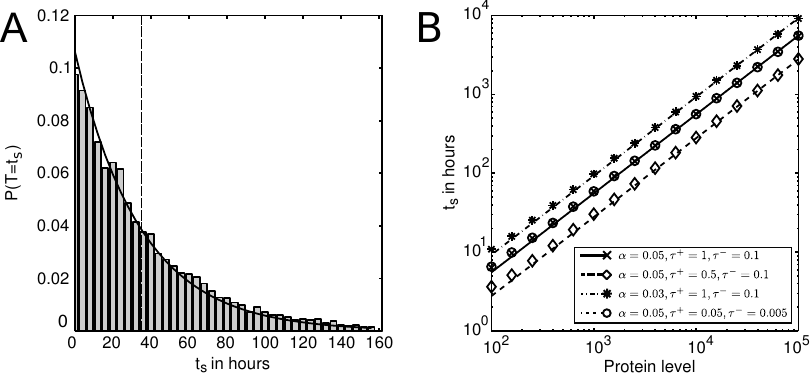}
  \caption{Residence time $t_s$ in the two-stage toggle switch. (A) The
    distribution for $t_s$ obtained by stochastic simulation is in good
    agreement with the geometric distribution derived from our
    mean-field approximation. The mean of the distribution is
    indicated by a dashed line. The protein decay rate was set to
    $\delta = 8 \cdot 10^{-4} s^{-1}$. (B) Mean residence time $t_s$ versus
    mean protein level $\bar{N}$ derived from stochastic simulation (symbols)
    and our analytical approximation (lines) for four different
    parameter settings. Note that the analytical approximations as well as the simulation results of the first and fourth parameter 
    set coincide.
    The exponent in the relation $t_s \propto
    ({\bar{N}_\playerOne})^\nu$ is $\nu = 1$, in accordance with equation \eqref{eq:3}. 
}
  \label{fig:geometricDist}
\end{figure}

We want to compare our analytical approximation with the residence
time derived from simulations. To that end, we have to infer the
dominating attractors from the simulated time courses.  
Recall that we can identify the dominating attractors via thresholds $\chi_\playerOne,\chi_\playerTwo$ at protein levels.
The residence time of attractor $S_\playerOne$ ($S_\playerTwo$) is estimated as the consecutive time in a 
trajectory where $N_\playerOne > \chi_\playerOne$ ($N_\playerTwo > \chi_\playerTwo$).
We compare the analytically derived geometric distribution for the residence
times (see Eq.~\eqref{geometric}) with numerical results by
simulating the switch with a given parameter set and estimating the residence times from 10000 stochastic simulations.
Fig.~\ref{fig:geometricDist}A shows
excellent agreement between the geometric distributed residence time
and the simulations for a protein degradation rate of $\delta = 8
\cdot 10^{-4} s^{-1}$. This legitimates the approximations and
assumptions made above for the parameter regime of rapid transcription
factor binding. From the analysis of the mean residence time for
different protein half-lives, we find again a good agreement between
the simulation and the approximation (see Fig.~\ref{fig:geometricDist}B).  
Moreover, the slope of the log-log curve of the simulation
is 1 -- confirming a linear dependence of the residence time from the
mean protein level.

With the result from Eq.~\eqref{geometric} we can compare the
mean residence time of different switch models.  First
we consider a gene expression model where transcription and
translation are condensed into a single protein synthesis reaction.
In analogy to the two-stage model of gene expression
\citep{shahrezaei08_proc-natl-acad-sci-u}, this can be called a
one-stage model of gene expression.  To achieve the same amount of
proteins at similar degradation rates, the synthesis rate in the
one-stage model needs to be larger compared to the transcription and
translation rates in the two-stage model.  The probability $q_t$ which
accounts for translation during mRNA lifetime can be set to 1, since
there is no mRNA stage and proteins are produced immediately.  The
binding probability $q_b$ remains unchanged.  However, because of the
increased synthesis rate, the probability $q_s$ of synthesis during the
unbound phase will be larger than in the two-stage model.  Therefore,
the mean residence time will be decreased in the one-stage model as
compared to the two-stage model, leading to more frequent attractor
changes. This finding is in accordance with the previously reported stabilizing effect of bursts in an exclusive switch \cite{Schultz2008}.

A second modification of the  switch includes not only mutual
inhibition but also autoactivation of both genes.  If the promoter
of the gene is unbound it will be transcribed with a small basal rate
$\kappa$.  If the promoter is bound by its own protein product the
gene will be transcribed with full rate $\alpha \gg \kappa$.  Repressor
bound promoters are inactive. For simplicity we assume that either
activators or repressors are bound but not both at the same time.
Note that in this case also the deterministic ODE model is bistable \citep{TwoGeneswitchStability}.
Considering the mean residence time in a two-stage  switch with
autoactivation, we find that the probability $q_s$ of mRNA synthesis
during the unbound phase is smaller than in the ordinary two-stage
model.  Since no activator is present in this attractor, mRNA has to
be transcribed with the small basal rate $\kappa$, making the
transcription more improbable.  The probability $q_t$ for translation
remains unchanged.  However, the probability $q_b$ of protein binding
to the antagonistic promoter is also decreased since this promoter is
occupied by the abundant activator most of the time.  Therefore,
repressor binding to this promoter requires an additional dissociation
reaction of the activator during repressor lifetime.  As both $q_s$
and $q_b$ are decreased the mean residence time in  switch
models with autoactivation will be strongly increased compared to the
ordinary two-stage model.

Summarizing, we find that the residence time is (i) geometrically
distributed, (ii) the mean of the distribution grows linearly with the
number of proteins for slow mRNA degradation, and (iii) both the
intermediate step of mRNA production and the autoactivation of transcription factors
increase the residence time.
\section{Discussion}
\subsection{Lineage priming}
We now discuss the implications of our findings in the context of cell differentiation driven by the toggle switch.
In previous studies \cite{roeder06_j-theor-biol,  huang07_dev-biol, wang10_biophys-j}, 
attractors where either one or the other player is dominating, thereby repressing the antagonist ($S_\playerOne$, $S_\playerTwo$) corresponded to committed cells.
We also find analogs for the intermediate states $S_\playerOne^*$ and $S_\playerTwo^*$. 
In these attractors the system has a strong preference towards one specific dominating attractor, but is not fully committed yet.
A similar behavior is known as lineage priming in stem cell biology \cite{Graf2008}. %
Two different studies \cite{MuellerSieburg2002,Chang2008a} showed that a population of stem and progenitor cells, 
respectively, can be divided into subpopulations that mainly give rise to only one of two possible cell types.
In our simple model this would correspond to stem cells that reside either in $S_\playerOne^*$ or $S_\playerTwo^*$.
These stem cells can still give rise to both cell fates but have a strong tendency towards one of them.

Remarkably only a two-stage probabilistic model of the toggle switch shows dynamics reminiscent of lineage priming.
Although a progenitor state exists in one-stage models of the toggle switch, cells in this state will move to either the one or the other committed state with equal probability.

\subsection{Residence time}
We find that the residence time in $S_\playerOne$ and $S_\playerTwo$, a key property of the
system, is geometrically distributed.  
Previous contributions \citep{bialek01_,warren05_j-phys-chem-b,barzel08_phys-rev-e-stat-nonl} 
focused only on the mean residence time and did not consider its underlying distribution.
What does a geometric distribution for the residence time imply for the differentiation process dependent on the state of a genetic switch?
To discuss this question, let us first
reason on how a differentiation decision could be established with the
toggle switch lined out in the previous sections.

We discriminate two scenarios for the differentiation of a cell: 
In the first scenario, the state of the switch completely determines the cell fate.
Starting in the progenitor attractors $S^*_\playerOne$ or $S^*_\playerTwo$, after a certain amount of time, 
the switch will move to a committed attractor. We assume that the high numbers of proteins of the dominating player will trigger
the differentiation program of the associated lineage and establish the mature cell type. 
However due to stochasticity, the switch will drop out of the committed attractors and the cell will not only lose the current lineage decision, 
but possibly even switch to the opposing cell fate.
In order to establish stable lineages in this scenario, the cell has to assure that the residence time of the switch is much longer 
than all relevant biological processes of the cell, especially cell lifetime. This guarantees that the cell will keep a lineage decision once it has obtained one.
Yet the geometric distribution of the residence time imposes difficulties in this scenario: 
Even if the mean residence time is high, short residence times will always be more probable than longer residence times.
The toggle switch could either be stabilized with the aforementioned autoactivation of the players, or with very high protein numbers
so that the geometric distribution flattens and transforms to an almost uniform distribution.
Both means would assure that only a very small percentage of a population of cells forgets its lineage decision during lifetime.

In the second scenario we assume that the cell gets locked in one fate by changing the 
shape of the underlying potential so that further transitions between attractors become less possible.
Such a change of the potential could for example be facilitated by chromatin changes,
as proposed by \citet{akashi03_blood}.
In the following, we assume
that only if one state dominates the other for a long enough fixation time
$t_d$, downstream genes necessary for the decision process are
activated (e.g. leading to chromatin remodeling), and the cell differentiates. Such a time depending property
could be implemented with low-pass filters (see
\cite{narula10_plos-comput-biol} for an example in hematopoietic stem cells) and would
allow for an integration of external signals (see
\cite{rieger09_science} for the instructive power of hematopoietic
cytokines).
In this scenario, the residence time determines when differentiation will occur:
The switch will constantly move into and out of the dominating attractors, until the residence time is finally long enough 
so that the dominating player can activate the downstream differentiation machinery.
Ignoring the time the system spends in the intermediate attractors and just summing up the residence times in $\SP$ and $\SG$
until a long enough residence time for differentiation occurs, we find that this time follows a geometric-like distribution (see Supporting Material 5).
Under this differentiation mechanism, most cells will differentiate very fast and only few cells will need longer.
Experiments that measure the time for single cells needed to go from the primed to the committed state 
(as an extension to the 2-day threshold reported by \citet{Heyworth2002} for GM-CTC cells) 
in order to support or reject these hypotheses remain to be done.

\subsection{Comparison to previous models}
Finally we discuss how our findings relate to previous studies on the toggle switch.
We found that the mean of the residence time distribution scales linearly with the
number of proteins in the system. 
The more proteins are present, the longer the average residence time in $S_\playerOne$ or $S_\playerTwo$. 
However, shorter residence times are still most probable due to the geometric distribution.
This holds for the one-stage, the two-stage, and the auto-activating scenario. 

This linear scaling differs from the exponential \citep{bialek01_} or near exponential
\citep{warren05_j-phys-chem-b} scaling described previously in the one-stage scenario. 
In contrast to our model, the model of \citet{warren05_j-phys-chem-b}
considers dimerization of the transcription factors, motivated by the fact that cooperative binding is 
necessary to achieve bistability in a deterministic framework \citep{chickarmane09_plos-comput-biol}.
We showed that, as soon as stochastic fluctuations are introduced, a system with multiple attractors is achieved that can act as a
proper switch with additional states of low co-expression. 
Including dimerization as a prerequisite for inhibition in a one-stage model will strongly increase the stability of the attractors $S_{\playerOne/\playerTwo}$ \citep{warren05_j-phys-chem-b}.
This is consistent with our findings:
Instead of requiring translation of one protein of the suppressed species, we now require this rare event to happen twice during a short time frame, which is much less probable.
However, the inclusion of dimerization will have less effect on the two-stage switch:
Since proteins are typically synthesized in bursts  (in our model the average burst size is $\beta/\gamma = 10$) and dimerization is a fast process \citep{warren05_j-phys-chem-b}, 
as soon as one burst occurs almost certainly a dimer is formed and can inhibit the currently dominating player.
Therefore the probability of leaving the attractors $S_{\playerOne/\playerTwo}$ is similar to a non-dimeric inhibition.

Contrary to our results, \citet{warren05_j-phys-chem-b} report that introduction of mRNAs reduces the stability of the switch. 
This discrepancy can be understood in the light of dimerization.
In their one-stage model dimerization is a key ingredient of stability, which is lost when introducing translational bursts (''shot noise'').
As we considered monomeric transcription factor binding, 
stability does not rely on dimerization. Therefore mRNAs increase the stability of the system, because they introduce additional conditions required for switching.

Due to these differences in the model it is hard to resolve the discrepancy between our linear and
the exponential scaling of residence time found by \citet{bialek01_} and \citet{warren05_j-phys-chem-b}.
However we want to emphasize that the theoretical results shown in \cite{warren05_j-phys-chem-b} only consider protein numbers up to 30. 
In this region our simulation results show slight deviations from the analytical linear dependence (Fig.~\ref{fig:geometricDist}).
At such low protein numbers the system does not only leave the dominating attractor according to the mechanism described in our results.
It is also likely that just due to fluctuations in the gene expression (not fluctuations in the promoter) the dominating attractor is left.
This mechanism operates only at very small protein numbers and its probability rapidly decreases with rising protein numbers.
Therefore our results do not contradict the findings of \citet{warren05_j-phys-chem-b}, but consider a different parameter regime with higher protein numbers.
Interestingly, the noise-driven attractor changes are also described by \citet{Kashiwagi2006} where the authors link this mechanism to the selection 
of a favorable, less noisy attractor in \textit{E. Coli} populations.

In another contribution, \citet{morelli08_biophys-j} use the forward flux sampling algorithm to assess the stability of a one-stage genetic toggle switch with dimeric transcription factor binding.
They find a similar mechanism of attractor flipping which is based on the synthesis of the suppressed species due to promoter fluctuations.
Using the forward flux sampling, they obtain estimates of the switching rate (the inverse of the mean residence time) 
for different amounts of fluctuations in DNA-protein interaction and dimerization.
\citet{morelli08_biophys-j} modulate the size of fluctuations at the promoter by varying the ratio of binding rate and synthesis rate, the adiabaticity parameter
$\omega = \tau^+ / \alpha$ ($\tau^-$ is adjusted to keep $\tau^+ / \tau^-$ constant). 
Small $\omega$ leads to strong fluctuations, whereas large $\omega$ reduces fluctuations.
They find that increasing $\omega$ decreases the average switching rate and therefore stabilizes the switch. This dependency vanishes for $\omega >5$, where the average switching rate remains constant.
The latter is in accordance with our results in Eq.~\eqref{eq:3}, where the mean residence time depends only on the ratio of $\tau^+$ and $\tau^-$, not on the absolute values 
and is therefore independent of $\omega$.
The dependency of the average switching rate for $\omega < 5$ is not predicted by Eq.~\eqref{geometric} and \eqref{eq:3}. 
It is also not visible in the stochastic simulations, where mean residence times of systems with $\omega =1$ and $\omega = 20$ coincide (Fig.~\ref{fig:geometricDist}).
The results of \citet{morelli08_biophys-j} were simulated for an average number of proteins $\bar{N}_\playerOne = \bar{N}_\playerTwo = 27$. 
As mentioned above, in regions of very small protein numbers the system might leave the dominating attractor by a mechanism not captured by Eq.~\eqref{geometric} and \eqref{eq:3}, 
probably causing the difference of the results of \citet{morelli08_biophys-j} and our results for small $\omega$.

\section*{Acknowledgments}
We thank Jan Krumsiek, Robert Schlicht, Timm Schroeder and Peter Swain for stimulating discussions 
and Christiane Dargatz and Dominik Wittmann
for careful reading and thoughtful feedback on drafts of this manuscript.
Thanks to Sabine Hug and Daniel Schmidl for their help concerning the stability analysis.
Moreover we acknowledge the comments of the unknown reviewers, who improved the quality of the manuscript.

This work was supported by the Helmholtz Alliance on Systems Biology (project 'CoReNe'), 
the European Research Council (starting grant 'LatentCauses'), and the German Science Foundation DFG (postdoc fellowship for C.M. and SPP 1356 'Pluripotency and Cellular Reprogramming').

\newpage

\clearpage

\end{document}


\title{Supporting Material\\Stability and multi-attractor dynamics of a toggle switch based on a
two-stage model of stochastic gene expression}

\author{\textbf{Michael~Strasser}\\
	Institute for Bioinformatics and Systems Biology, \\
	Helmholtz Zentrum M\"unchen\\
	German Research Center for Environmental Health, Germany
	\and \textbf{Fabian~J.~Theis} \\
	Institute for Bioinformatics and Systems Biology,\\
	Helmholtz Zentrum M\"unchen\\
	German Research Center for Environmental Health, Germany\\
	{\small and}\\
	Institute for Mathematical Sciences,\\
        Technische Universit\"at M\"unchen\\
	Garching, Germany
	\and \textbf{Carsten~Marr}\thanks{
           Corresponding author.  Address:
           Institute for Bioinformatics and Systems Biology,
	   Helmholtz Zentrum M\"unchen,
	   Ingolstaedter Landstrasse 1,
	   85764 Neuherberg, Germany,
	   Tel.: +49 (0)89 3187 3642, Fax: +49 (0)89 3187 3585, Email:~carsten.marr@helmholtz-muenchen.de}
	\thanks{Present address: Centre for Systems Biology at Edinburgh, University of Edinburgh, Edinburgh EH9 3JD, UK}
	\\
	Institute for Bioinformatics and Systems Biology, \\
	Helmholtz Zentrum M\"unchen\\
	German Research Center for Environmental Health, Germany}

\date{\today}
\maketitle

\section{Steady state solution of the deterministic two-stage\\toggle switch}\label{app:twoStageODe}
Using symmetric parameters $\alpha = \alpha_\playerOne = \alpha_\playerTwo$, $\beta = \beta_\playerOne = \beta_\playerTwo$, 
$\gamma = \gamma_\playerOne = \gamma_\playerTwo$, $\delta = \delta_\playerOne = \delta_\playerTwo$, 
$\tau^+ = \tau^+_\playerOne = \tau^+_\playerTwo$ and $\tau^- = \tau^-_\playerOne = \tau^-_\playerTwo$ 
results in two following steady state solutions of
equations (7), (8) and (9) in the main text:
\begin{align}
\nMP^{(1)} &= \nMG^{(1)} = -\frac{\delta \tau ^- }{2 \beta  \tau ^+} \left(1  -\eta\right)\label{app:odeSol1}\\
\nP^{(1)} &= \nG^{(1)} = -\frac{\tau ^-}{2 \tau ^+}\left(1 -\eta\right)\label{app:odeSol2}\\
\nDP^{(1)} &= \nDG^{(1)} = \frac{2}{1+\eta}\label{app:odeSol3}
\end{align}

\begin{align}
\nMP^{(2)} &= \nMG^{(2)} = -\frac{\delta \tau ^- }{2 \beta  \tau ^+} \left(1  +\eta\right)\\
\nP^{(2)} &= \nG^{(2)} = -\frac{\tau ^-}{2 \tau ^+}\left(1 +\eta\right)\\
\nDP^{(2)} &= \nDG^{(2)} = \frac{2}{1-\eta}
\end{align}
with $\eta = \sqrt{\frac{4 \alpha  \beta  \tau ^+}{\gamma \delta \tau^-}+1}$.
The first solution is positive, the second is negative (given all parameters are positive).
Only the positive solution is of interest in biological systems.
The monostability of this system is due to 
the monomeric protein binding. As soon as dimeric binding is introduced, the system becomes multistable, similar 
to the existing one-stage deterministic models \cite{roeder06_j-theor-biol,  huang07_dev-biol,chickarmane09_plos-comput-biol}.

Note that for small $\tau^+$ equations \eqref{app:odeSol1} and \eqref{app:odeSol2} reduce to the steady state solution of a 
simple two stage expression model 
\begin{align*}
\nP^{(1)} &= \nG^{(1)} =\frac{\alpha \beta}{\gamma \delta}\\ 
\nMP^{(1)} &= \nMG^{(1)}= \frac{\alpha}{\gamma} \; ,
\end{align*}
because \[\eta \approx  1+ \frac{2\alpha \beta \tau^+}{\gamma \delta \tau^-}\] for small $\tau^+$  through the Taylor approximation \[(1+x)^n \approx 1+nx \text{ for } |x|\ll1 \;.\]
This is expected since setting $\tau^+$ to 0 removes the interaction between both players, which will then evolve independently according to a two-stage expression model.

For decreasing $\tau^-$ the solution of the system approaches the origin 
\begin{align*}
 \nP^{(1)} &= \nG^{(1)} = 0\\
\nMP^{(1)} &= \nMG^{(1)} = 0
\end{align*}
because $\eta \approx \frac{\text{const}}{\sqrt{\tau^-}}$ and therefore the right hand sides of equations \eqref{app:odeSol1} and \eqref{app:odeSol2} reduce to
 $\text{const}\cdot \frac{\tau^-}{\sqrt{\tau^-}} $. For decreasing $\tau^-$ this term will approach 0.
Hence, if proteins never unbind the promoter, the system will be locked forever yielding 0 protein and mRNA levels in steady state.

We now assess the stability of the positive solution \eqref{app:odeSol1} - \eqref{app:odeSol3} using standard linear stability analysis. 
To reduce the complexity of our system for the stability analysis, we apply a quasi steady state approximation to the DNA binding/dissociation process ($\dot{\nDP} = \dot{\nDG} = 0$), 
reducing the dimensionality of our system to four equations:
\begin{align*}
    \ddt \nMP &= \alpha  \psi(\nG) - \gamma  \nMP\\
      \ddt \nMG  &= \alpha  \psi(\nP) - \gamma  \nMG\\
    \ddt \nP &= \beta  \nMP - \delta  \nP 
     + \tau^- (1-\psi(\nP)) - \tau^+  \psi(\nP)  \nP\\ 
      \ddt \nG  &= \beta  \nMG - \delta  \nG 
     +\tau^- (1-\psi(\nG)) - \tau^+  \psi(\nG)  \nG\;,
\end{align*}
with $\psi(x) = \frac{\tau^-}{\tau^- + x \cdot \tau^+}$.
The reduced system has the positive steady state solution
\begin{align}
 \nMP^{(ss)} &= \nMG^{(ss)} = -\frac{\delta \tau ^- }{2 \beta  \tau ^+} \left(1  -\eta\right)\label{eq:reducedodeSol1}\\
  \nP^{(ss)} &= \nG^{(ss)} = -\frac{\tau ^-}{2 \tau ^+}\left(1 -\eta\right)\label{eq:reducedodeSol2}
\end{align}
with $\eta = \sqrt{\frac{4 \alpha  \beta  \tau ^+}{\gamma \delta \tau^-}+1}$.
Notice that this is the same as the solution for mRNA and protein of the full system (Eq.~\eqref{app:odeSol1} and Eq.~\eqref{app:odeSol2}).
We calculate the Jacobian matrix of the reduced system as 
\begin{align*}
  J = \begin{pmatrix}
  -\gamma & 0       & 0  & \frac{\alpha \tau^+}{\tau^-} \cdot \psi^2(\nG)\\
  0       & -\gamma & \frac{\alpha \tau^+}{\tau^-} \cdot \psi^2(\nP)   & 0\\
  \beta   & 0       & -\delta + \tau^+ \psi^2(\nP) [1+ \tau^+ \nP- \psi^{-1}(\nP)]  & 0\\
  0       & \beta  & 0 & -\delta + \tau^+ \psi^2(\nG) [1+ \tau^+ \nG- \psi^{-1}(\nG)]
 \end{pmatrix}\;.
\end{align*}

\begin{figure}[h]
  \centering
  \includegraphics[width=1\textwidth]{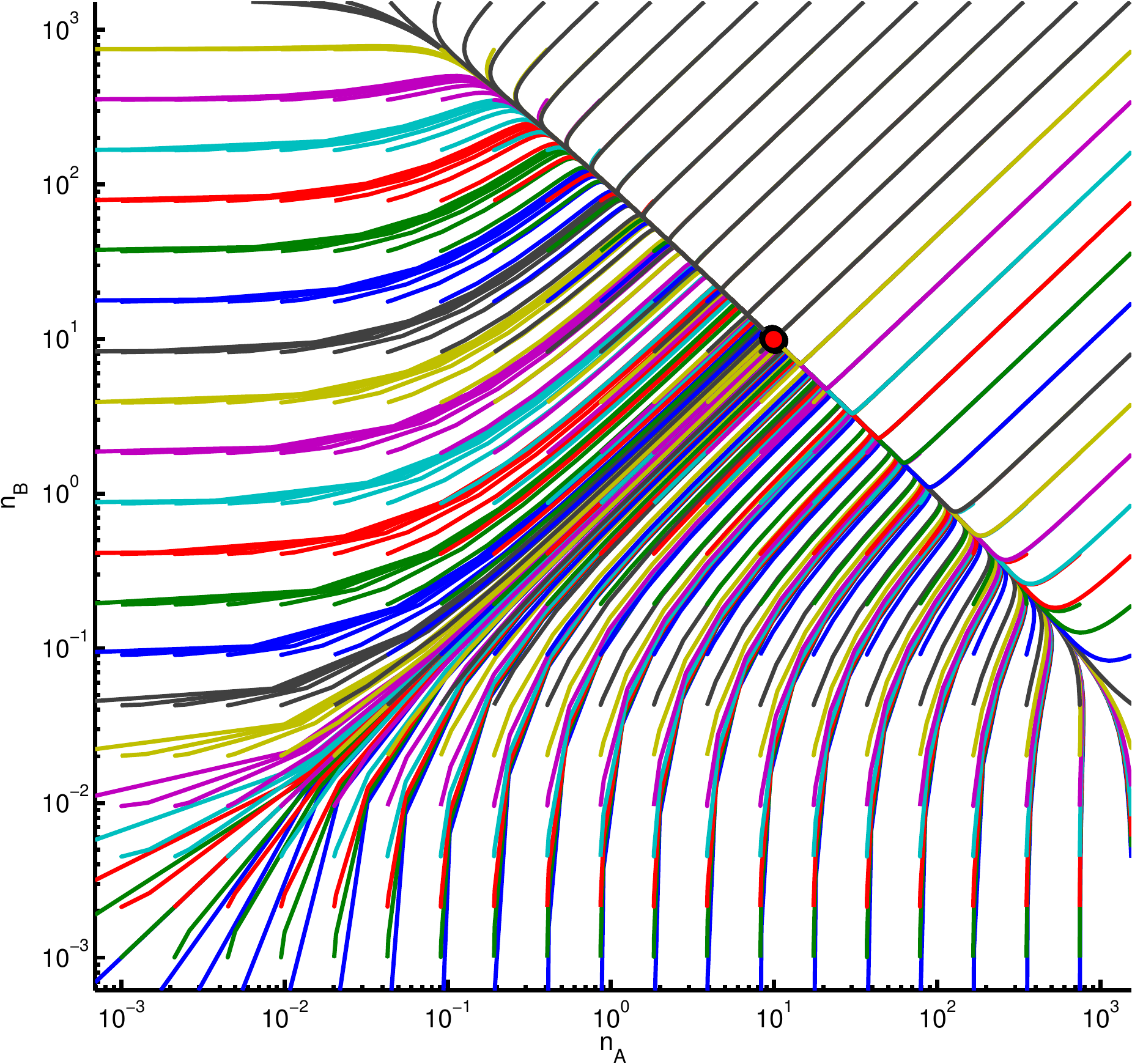}
  \caption{Phase portrait of the deterministic two-stage toggle switch model projected onto the $N_\playerOne$ and $N_\playerTwo$ dimensions.
The same parameters were used as in the probabilistic model of Fig.\;2 and  Fig.\;3.  
The ODE was solved for different initial protein concentrations $\nP,\nG$ and the mRNA concentrations where set to $\nMP = 0.01 \cdot \nP$ and $\nMG = 0.01 \cdot \nG$, 
  resembling the fact that for the given parameters each mRNA will correspond to approximately 100 proteins.
Trajectories are arbitrarily colored to improve readability. Note that trajectories can intersect as this in only a projection of the full system.
Inspection of the phase portrait reveals the single steady state (red dot) as predicted by Eq.\;\eqref{eq:reducedodeSol1} - \eqref{eq:reducedodeSol2}. No limit cycles are visible in the phase portrait.
}
  \label{fig:ODESOLUTION}
\end{figure}

We evaluate the Jacobian at the steady state solution of the reduced system (Eq.\;\eqref{eq:reducedodeSol1}-\eqref{eq:reducedodeSol2})
and use the Hurwitz criterion to verify that all its eigenvalues have negative real part. 
We conclude that the system has one stable positive fixed point but we cannot analytically exclude the existence of limit cycles.
However, inspection of the system's phase portrait (see Fig.\;\ref{fig:ODESOLUTION}) indicates that no limit cycles exist.

\label{sec:appendix}

\section{Master equation}
\label{sec:master-equation}

The stochastic model of the two-stage  switch is completely
defined by the master equation. We define $\mathcal{P}_{ij}(\discreteMP,\discreteMG,\discreteP,\discreteG,t)$
as the probability at time $t$ to have $\discreteMP$ copies of $\mRNAP$, $\discreteMG$ copies of
$\mRNAG$, $\discreteP$ copies of $\proteinP$, $\discreteG$ copies of $\proteinG$,
and the corresponding promoter configuration $ij$ where $0$ $(1) $
means an unbound (bound) promoter.  This results in the following
master equation that is split up into four coupled equations,
corresponding to the four promoter states:

\begin{align*}
\ddt \mathcal{P}_{00}(\discreteMP,\discreteMG,\discreteP,\discreteG,t) &= \tau^-_\playerOne
   E^-_{\discreteP} \mathcal{P}_{01}(\discreteMP,\discreteMG,\discreteP,\discreteG,t) + \tau^-_\playerTwo
    E^-_{\discreteG} \mathcal{P}_{10}(\discreteMP,\discreteMG,\discreteP,\discreteG,t)\\
   &+[-\tau^+_\playerTwo \discreteG- \tau^+_\playerOne \discreteP +
   \alpha_\playerOne (E_\discreteMP^- -1) + \alpha_\playerTwo (E_\discreteMG^- -1)
   + \gamma_\playerOne  (E_\discreteMP^+ -1)\cdot \discreteMP + \gamma_\playerTwo 
(E_\discreteMG^+ -1)\cdot \discreteMG\\
   &\quad+\beta_\playerOne (E^-_\discreteP-1)\cdot \discreteMP + \beta_\playerTwo
   (E_\discreteG^- -1)\cdot  \discreteMG + \delta_\playerOne (E^+_\discreteP -1)\cdot \discreteP+
\delta_\playerTwo
   (E^+_\discreteG -1)\cdot \discreteG] \\
   &  \cdot \mathcal{P}_{00}(\discreteMP,\discreteMG,\discreteP,\discreteG,t)\\
 \ddt \mathcal{P}_{11}(\discreteMP,\discreteMG,\discreteP,\discreteG,t) &=  \tau^+_\playerOne 
E^+_{\discreteP} \discreteP  \mathcal{P}_{10}(\discreteMP,\discreteMG,\discreteP,\discreteG,t) + \tau^+_\playerTwo 
E^+_{\discreteG} \discreteG  \mathcal{P}_{01}(\discreteMP,\discreteMG,\discreteP,\discreteG,t)]\\
   &+[-\tau^-_\playerOne - \tau^-_\playerTwo + \gamma_\playerOne 
(E_\discreteMP^+ -1)\cdot \discreteMP+ \gamma_\playerTwo  (E_\discreteMG^+ -1)\cdot \discreteMG\\
   &\quad+\beta_\playerOne (E^-_\discreteP-1)\cdot \discreteMP + \beta_\playerTwo 
(E_\discreteG^- -1)\cdot  \discreteMG + \delta_\playerOne  (E^+_\discreteP -1)\cdot \discreteP+
\delta_\playerTwo  (E^+_\discreteG -1) \cdot \discreteG]\\
   & \cdot \mathcal{P}_{11}(\discreteMP,\discreteMG,\discreteP,\discreteG,t)\\
 \ddt \mathcal{P}_{10}(\discreteMP,\discreteMG,\discreteP,\discreteG,t) &=  \tau^-_\playerOne E^-_{\discreteP}
\mathcal{P}_{11}(\discreteMP,\discreteMG,\discreteP,\discreteG,t) + \tau^+_\playerTwo E^+_{\discreteG} \discreteG
\mathcal{P}_{00}(\discreteMP,\discreteMG,\discreteP,\discreteG,t)\\
   &+[-\tau^-_\playerTwo - \tau^+_\playerOne \discreteP
   +\alpha_\playerTwo (E_\discreteMG^- -1)  + \gamma_\playerOne  (E_\discreteMP^+
-1)\cdot \discreteMP+ \gamma_\playerTwo  (E_\discreteMG^+ -1)\cdot \discreteMG\\
   &\quad+\beta_\playerOne (E^-_\discreteP-1)\cdot \discreteMP + \beta_\playerTwo 
(E_\discreteG^- -1)\cdot  \discreteMG +\delta_\playerOne  (E^+_\discreteP -1)\cdot \discreteP+
\delta_\playerTwo  (E^+_\discreteG -1)\cdot \discreteG]\\
   & \cdot \mathcal{P}_{10}(\discreteMP,\discreteMG,\discreteP,\discreteG,t)\\
 \ddt \mathcal{P}_{01}(\discreteMP,\discreteMG,\discreteP,\discreteG,t) &=  \tau^+_\playerOne E^+_{\discreteP}
\discreteP \mathcal{P}_{00}(\discreteMP,\discreteMG,\discreteP,\discreteG,t) + \tau^-_\playerTwo E^-_{\discreteG}
\mathcal{P}_{11}(\discreteMP,\discreteMG,\discreteP,\discreteG,t)\\
   &+[-\tau^+_\playerTwo \discreteG - \tau^-_\playerOne
   +\alpha_\playerOne (E_\discreteMP^- -1)  + \gamma_\playerOne  (E_\discreteMP^+
-1)\cdot \discreteMP+ \gamma_\playerTwo  (E_\discreteMG^+ -1)\cdot \discreteMG\\
   &\quad+\beta_\playerOne (E^-_\discreteP-1)\cdot \discreteMP + \beta_\playerTwo 
(E_\discreteG^- -1)\cdot  \discreteMG + \delta_\playerOne  (E^+_\discreteP -1)\cdot \discreteP+
\delta_\playerTwo  (E^+_\discreteG -1)\cdot \discreteG]\\
   & \cdot 
\mathcal{P}_{01}(\discreteMP,\discreteMG,\discreteP,\discreteG,t)\;.
\end{align*}
The shift operators $E^+_x$ and $E^-_x$ increase or decrease the
function argument $x$ by one, $ E^\pm_x f(x) = f(x\pm1).$ To our
knowledge no results have yet been published on the solution of
stochastic two-stage switches.

\section{Rates}
\label{sec:rates}

\begin{table}[b]
  \begin{center}
    \begin{tabular}{p{0.3\textwidth}l}
      Reaction & Parameter value\\\hline
      Transcription  $\alpha$ & $0.05 s^{-1}$\\
      Translation  $\beta$ & $0.05\ s^{-1}\text{mRNA}^{-1}$\\
      mRNA degradation  $\gamma$ & $0.005\ s^{-1}$ \\
      Protein degradation  $\delta$ & $5\cdot10^{-3} $ to 
        $5\cdot10^{-6} s^{-1}$\\

      DNA binding  $\tau^+ $ & $ 1\ s^{-1}\text{Protein}^{-1}$\\
      DNA dissociation $\tau^- \; $ & $0.1\ s^{-1}$\\
    \end{tabular}
  \end{center}
  \caption{Parameters of the  switch model used throughout this
    work. Protein degradation is chosen according to the desired
    protein level $n$. If not mentioned otherwise, all simulations and
    plots are based on this set of parameters.}
  \label{tab:paras}
\end{table}

First we derive upper boundaries for the transcription and translation
rates.  Transcription of DNA into mRNA is accomplished by the
RNA-polymerase.  One polymerase can process about 10-20 nucleotides
(nt) per second in eucaryotes
\cite{alberts02_,Dahlberg1991,Singh2007}. As described by
\citet{alberts02_} the newly elongated RNA fragment is immediately
released from the DNA, which enables other polymerases to follow up
even before the first mRNA has been completed.  The distance $d$
between polymerases is estimated to be around 100 nt
\cite{Kennell1977}.  The rate of transcription is independent of the
sequence length $l$, since the longer the gene, the more polymerases
can process it in parallel.  Altogether we find the maximal
transcription rate $\alpha$ by dividing the speed of transcription $v$ with the
sequence length $l$, multiplied with the number of transcribing
polymerases,
\begin{equation*}
  \alpha = \frac{v}{l} \cdot \frac{n}{d} \approx \frac{10 \;
  \mathrm{nt}/\mathrm{s}}{l} \cdot \frac{l}{100 \; \mathrm{nt}} = 0.1 s^{-1}
\end{equation*}
required that enough polymerases and nucleotides are present.
  
The maximal translation rate can be inferred in a similar way:
Ribosomes, large complexes of proteins and rRNAs that translate mRNA
into polypeptides, proceed with a speed $v$ of 2 codons ($=6$ nt  $=2$
amino acids) per second in eucaryotes \cite{alberts02_}.
One mRNA can be processed by many ribosomes (polyribosomes) at the
same time \cite{alberts02_}.  The average space between two
ribosomes is 80 nt or $\approx$ 27 amino acids (AA). 
\cite{alberts02_}.  Therefore the overall translation rate for
an mRNA of length $n$ is
\begin{equation*}
  \beta \approx \frac{2 \textrm{AA}/s}{l} \frac{l}{27 \textrm{AA}} = 0.074
  s^-1 \;,
\end{equation*}
again independent of the mRNA length $l$.  This corresponds to the
maximally possible translation rate.  The actual rate will be smaller
when not enough ribosomes or other involved molecules (tRNA, amino
acids) are present. We estimate the minimal translation time as
$1/0.074s^{-1} = 13.5s$, which is in good agreement with literature,
where the time needed for one translation is said to be between 20
seconds and several minutes \cite{alberts02_}.
Notably, these transcription and translation rates only provide rough
estimates of the relevant timescale. Throughout the manuscript, 
we use a transcription and translation rate of $\alpha = \beta = 0.05\
s^{-1}$, corresponding to an average time of 20 seconds per product,
which seems to be reasonable in the context of the above
considerations. The fact that both rates are equal is not expected to
have influences on the results.
  
Interactions between proteins and DNA are mediated by specific regions
of the proteins, called DNA-binding domains, which on the one hand can
recognize specific DNA sequences and on the other hand maintain the
interaction between DNA and protein.  Zinc Fingers, Leucine Zippers or
Helix-Turn-Helix motifs are prominent examples of DNA binding domains
\cite{alberts02_}.  The binding between DNA and protein is
maintained by hydrogen bonds, ionic bonds, and hydrophobic
interactions.  Single interactions are weak, but as many bonds are
formed, the binding between DNA and protein becomes stronger.  The
binding rates are very fast compared to transcription and translation
processes and according to \citet{alon07_} in the range of $1s^{-1} 
\textrm{Protein}^{-1}$
in \textit{Eschericia coli}.  The unbinding rate depends on the strength of
the interaction and is assumed to be 10 times smaller ($0.1s^{-1}$) in
our model, leading to strong binding of the protein to the DNA.  All
reaction rates of the models used in this work are summarized in Table
\ref{tab:paras}.

As we showed above, the transcription and translation rates have upper
bounds.  The only way for a cellular system to further increase the
abundance of proteins is to modulate the degradation rates of mRNA or
proteins, giving longer lifetimes to mRNA and proteins.  Thus, during
this work we manipulate the degradations rates to adjust the system's
protein level to a desired steady state.  Notably, following the
report of \citet{warren06_proc-natl-acad-sci-u}, mRNA levels are set
to 10 by adjusting the decay rate.

\section{Attractor boundaries}
\label{sec:regime-inference}
In this section we assume symmetric parameters for both players ($\alpha = \alpha_\playerOne = \alpha_\playerTwo, \ldots$)
for simplicity.
\subsection{$S_\playerOne$ and $S_\playerTwo$}
We approximate the protein number
distribution in the attractors $\SP$ and $\SG$ using results from
\citet{thattai01_proc-natl-acad-sci-u}, who showed that for a simple
two-state expression model, the mean and variance of protein numbers
obey
\begin{equation*}
  \bar{N} = \frac{\alpha \beta}{\gamma \delta} \;\;
  \mathrm{and}  \;\;
  \sigma^2 = \frac{\beta^2  \alpha}{\gamma^2 \delta +
    \delta^2\gamma}\;,
\end{equation*}
respectively.  Thereby, we assume that in the dominating attractors, the
presence of the antagonist can be neglected due to its marginal
transcription.
We define the boundary $\chi_x$ of attractor $S_x$ using a normal
approximation of the dominating protein's distribution as
\begin{align}
  \chi_x =\bar{N}_x - Z_q \cdot
  \sigma_x\label{eq:regimeBoundsTwoStage}\;,
\end{align}
where $Z_q$ is the $q$\% quantile of the standard normal distribution
of the protein number with mean $\langle n \rangle_x$ and standard deviation
$\sigma_x$. Using $q = 0.1$ throughout our study assures that 99.9\% percent
probability mass of the distribution lies beyond the lower boundary.
Therefore we are certain to capture all relevant protein numbers
belonging to $\SP$ and $\SG$. Using these boundaries, we can
define the attractors $\SP$, $\SG$, and $S_0$ accordingly:
\begin{align*}
  \SP &= \{s \in S | \discreteP > \chi_\playerOne \wedge \discreteG < \chi_\playerTwo\}\\\notag
  \SG &= \{s \in S | \discreteP < \chi_\playerOne \wedge \discreteG > \chi_\playerTwo\}\\\notag
  S_0 &= \{s \in S | \discreteP < \chi_\playerOne \wedge \discreteG < \chi_\playerTwo\} \;.\notag
\end{align*}

\subsection{$S_\playerOne^*$ and $S_\playerTwo^*$}
In the $S_\playerOne^*$/$S_\playerTwo^*$ states, one protein of the 'repressed' species is present, partially blocking the promoter of the dominating player.
Therefore the dominating player is not synthesized with full rate and its protein level will be smaller:

The promoter of the dominating player $\playerOne$ is free only $100 \cdot \frac{\tau^-}{\tau^+ + \tau^-}$ \% of the time.
Therefore the average number of $\mRNAP$ in this attractor is $m = \frac{\tau^-}{\tau^+ + \tau^-} \cdot \frac{\alpha }{\gamma}$ (compared to $\frac{\alpha}{\gamma}$ at full synthesis).
Consequently the average number of $\proteinP$ in the state $S_\playerOne^*$ will be 
\begin{align}
\bar{N}_\playerOne = m \cdot \frac{ \beta}{\delta} =  \frac{\tau^-}{\tau^- + \tau^+} \cdot \frac{\alpha \beta}{\gamma \delta}\label{eq:Starregimes}\;.
\end{align}
Based on this finding one can construct boundaries in a similar way to the $\SP$/$\SG$ attractors.

\section{Implications for Differentiation}
\label{sec:app:diff}

In the following, we assume
that only if one player dominates the other one for a sufficiently long time
$t_d$, downstream genes necessary for the decision process are
activated (e.g. leading to chromatin remodeling), and the cell differentiates.
We are thus interested in the
probability that the residence time $T$ in a dominating attractor is longer
than $t_d$:
\[\omega = P(T\ge t_d)\;.\]
This can easily be evaluated using the previous results on the
residence time.
Each time the switch flips, the system has the chance
to settle in a sufficiently long residence time.  
What is the probability that after switching $m$ times the residence time is greater than $t_d$? 
We therefore define a random variable $M$ representing the number of switching events until the residence time is greater than $t_d$
and find that it follows a geometric distribution with
parameter $\omega$:
\[P(M=m) = \omega(1-\omega)^{m-1}\;.\] 
This distribution tells us how likely it is to find a cell differentiated after $m$ switching events.
It is non-trivial to directly relate $m$ to the time until differentiation (further denoted by the random variable $T_p$) since this time
is composed of the actual realizations of the residence time and the time the system spends in the intermediate attractors.
As no analytic expression of the time in the intermediate attractors $\SP^*$ and $\SG^*$ is available, 
we do not take into account the time spent in $\SP^*$ and $\SG^*$ but only the residence time of the system in $\SP$ and $\SG$.

\begin{figure}[h]
  \centering
  \includegraphics[width=1\textwidth]{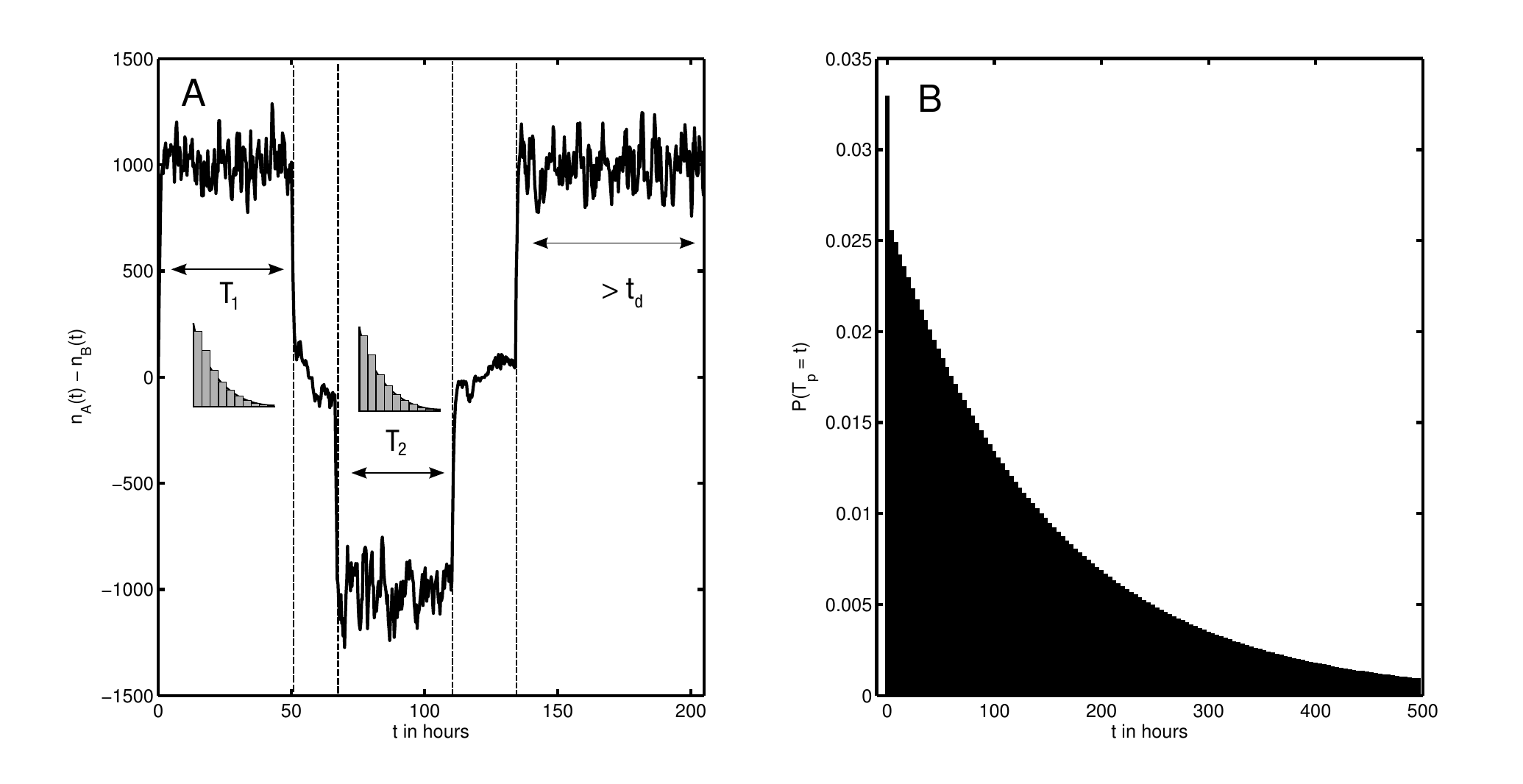}
  \caption{A: Scheme of the differentiation mechanism. The system changes the dominating attractor twice before it settles into $S_A$ for a time greater than $t_d$, which induces differentiation. 
	  The residence times in $S_\playerOne$ and $S_\playerTwo$ follow a (truncated) geometric distribution. The overall time until differentiation is therefore (neglecting the intermediate attractors) 
	  the sum of these geometric random variables.
	  B: Distribution of the differentiation time $T_p$ for a residence time of $1h$ and $t_d = 5h$. 
      Small differentiation times are more probable than higher bigger differentiation times. 
     Note the strong peak at $t=0$ indicating that it is most probable for a cell to differentiate immediately .
	  }
  \label{fig:differentiation}
\end{figure}

Now the time until differentiation only depends on the $m-1$ independent realizations of the residence time (see Fig.~\ref{fig:differentiation}A): 
\begin{align*}
T_p = \sum_i^{m-1}T_i \label{eq:tp}\;.
\end{align*}
The system flips the attractors $m-1$ times and resides here for a time given by $T_i$ until it stays in one dominating attractor for a time longer than $t_d$.
$T_p$ is a sum of random variables $T_i$ but also the number of summands varies according to the random variable $M$.

Another obstacle is the fact that the $T_i$ are not exact geometric random variables, but truncated above $t_d$.
For simplicity we do not make this distinction here. This is justified for small $\omega$, because $\omega$ determines the fraction of probability mass that is truncated.

It can be shown that the sum $Z$ of $n$ geometric random variables $T_i$ with common parameter $p$ follows a negative binomial distribution:
      \[Z = \sum_{i=1}^{n}T_i \sim NB(n,p)\]
\[P(Z=z) =\binom{z+n-1}{z}(1-p)^n p^z\;.\]
Since in our case $n$ is drawn from a distribution itself, we have to consider 
the compound distribution \citep{Baileybook} of $M$ and $Z$ 
to ultimately  obtain the probability density function for $T_p$,
 which is a combination of how many switching events take place 
(represented by $M$) and the actual time the system spends in the dominating attractors (represented by the $T_i$):

\begin{align*}
P(T_p=t) &= \sum_{m=1}^{\infty} P(M=m)P(T_p=t|M=m)\\
         &= \sum_{m=1}^{\infty} \omega (1-\omega)^{m-1} \binom{t+(m-1)-1}{t}(1-p)^{m-1} p^t\\
&= \begin{cases} -(1 - p)^t p [1 + p (-1 + \omega)]^{-1 - t} (-1 + \omega) \omega & t >0 \\ 
    \frac{\omega}{1 + p (-1 + \omega)} & t=0 \end{cases}\;.
\end{align*}
(A special case has to be considered for $m=1$, where the system immediately starts differentiating. Here one has to define  $P(T_p=0|M=1)= \omega$ and 
 $P(T_p>0|M=1)= 0$, since the negative binomial distribution $NB(n,p)$ is meaningless for $n=0$).

Fig.~\ref{fig:differentiation} B shows the distribution of $T_p$ for a mean residence time $t_s = 1h$ and a threshold of $t_d = 5h$, 
meaning that the system has to reside for more than $5h$ in the same attractor to promote differentiation.
Immediate differentiation is most probable ($t=0$) and the probability for higher differentiation times decreases quickly.
Therefore the majority of cells will differentiate rapidly and only few cells will need longer time to differentiate.
This is somewhat counterintuitive, since one would expect that it takes a long time until a sufficiently large residence time $> t_d$ has been accomplished to promote differentiation 
(since longer residence times are less likely than shorter ones).
However, due to the ''odd'' nature of the geometric distribution -- where the probability for success is always the highest on the first try -- 
the system has the highest chance to reach a sufficiently long residence time on the first try (reflected by the random variable $M$).
Even if the system has to change attractors more often, the differentiation time will be short since the residence time in each attractor 
is most likely very small (since it is itself a geometric random variable), 
contributing only little to the overall differentiation time.

\newpage
\newcommand{\force}{outflux\;}
\section{Supplementary Figures}
\newpage
\begin{figure}[h]
  \centering
  \includegraphics[width=1\textwidth]{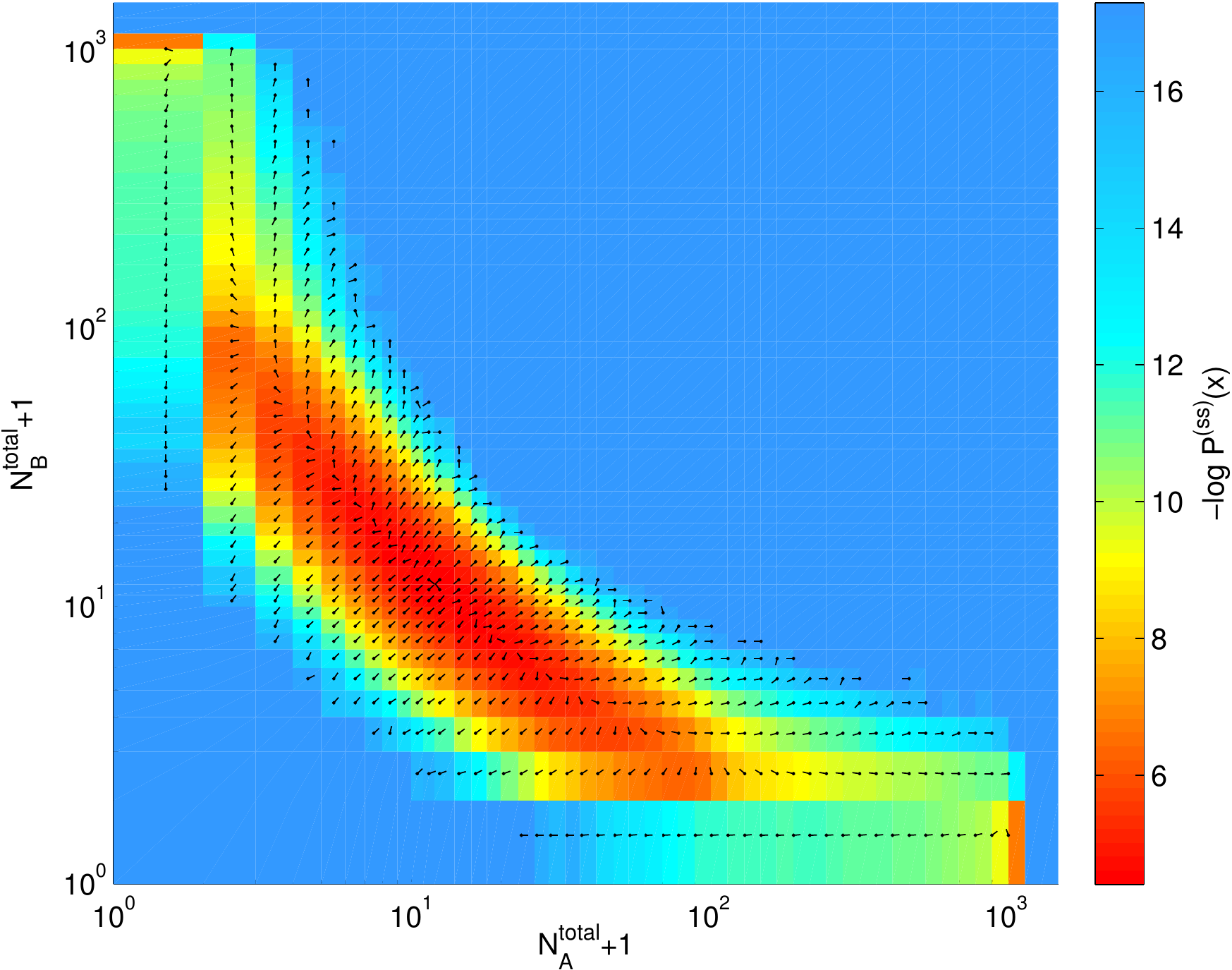}
  \caption{Quasi-potential of a toggle switch based on a one-stage expression model projected onto the $N_\playerOne^{\textrm{total}}$ and $N_\playerTwo^{\textrm{total}}$ dimensions.
	  Notice that both axis are on logarithmic scale and are shifted by 1 in order to include $N_\playerOne^{\textrm{total}}=0$ and $N_\playerTwo^{\textrm{total}}=0$. 
	  Therefore the lowest row in the plot corresponds to the case $N_\playerTwo^{\textrm{total}}=0$.
	  The quasi-potential $U(x) =  -\log {\cal P}^{(ss)}(x)$ is color coded where red areas reflect minima of the landscape. Contrary to the two-stage toggle switch model only three attractors are present.
	  The vectors of the \force at each point in state space are drawn as lines (circles correspond the the origin of the vector).
	  Vectors are normalized and therefore show only the direction, not the magnitude of the field. 
	  Parameters were set in accordance to the two-stage switch: 
	Protein synthesis $\alpha = 0.5 \;s^{-1}$, Protein degradation $\gamma = 5\cdot 10^{-4}\;s^{-1}$, DNA binding $\tau^+ = 1\; \text{Protein}^{-1}s^{-1}$, DNA dissociation $\tau^- = 0.1\;s^{-1}$.}
  \label{fig:oneStageLandscape}
\end{figure}
\newpage
\begin{figure}[h]
  \centering
  \includegraphics[width=1\textwidth]{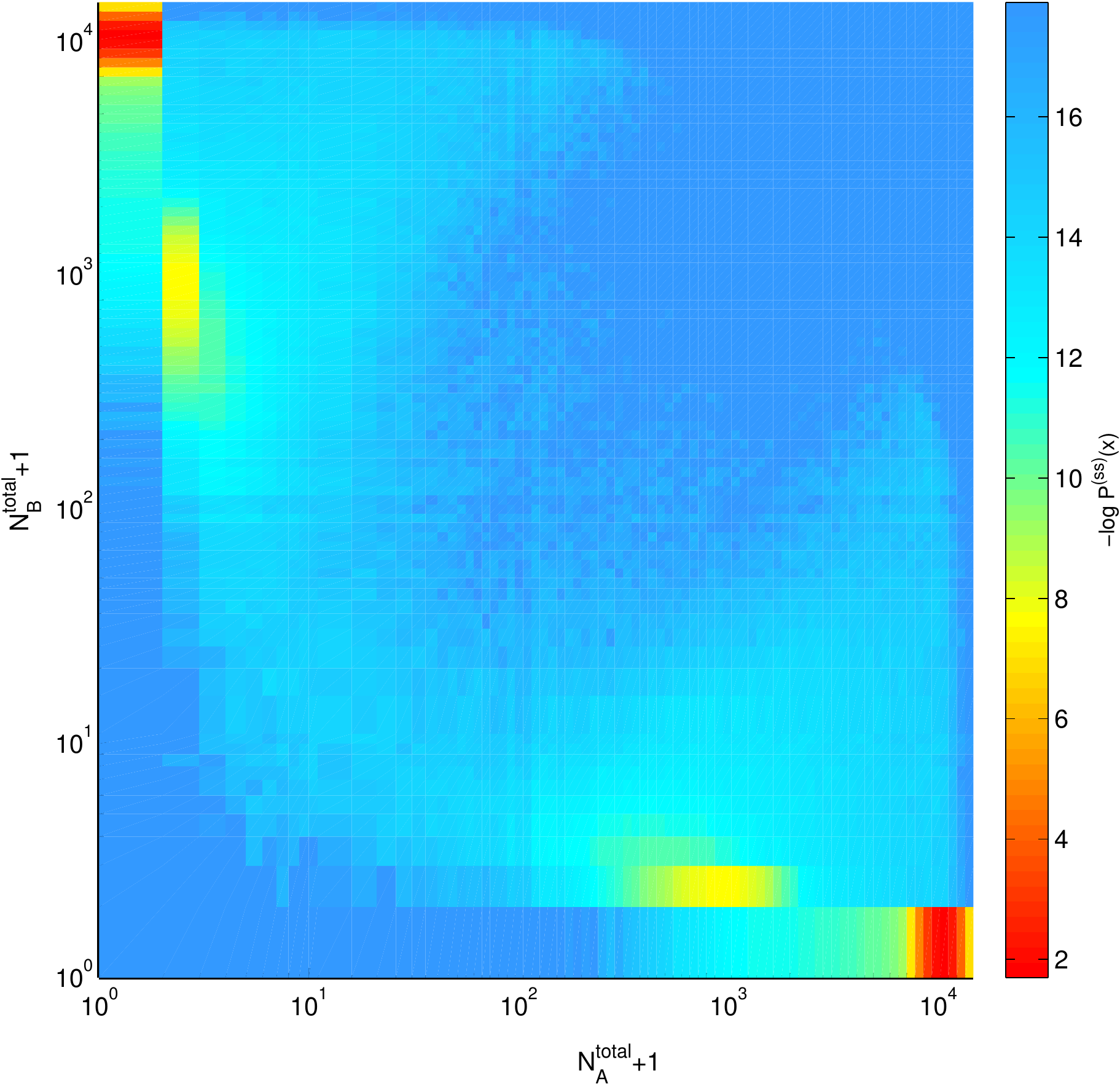}
  \caption{Quasi-potential of a two-stage toggle switch projected onto the $N_\playerOne^{\textrm{total}}$ and $N_\playerTwo^{\textrm{total}}$ dimensions using a different parameter set:
   Motivated by \citet{Hensold1996} we set the mRNA degradation rate to $\gamma = 3.6\cdot 10^{-5} s^{-1}$ and adjust the transcription rate $\alpha = 3.6\cdot 10^{-4} s^{-1}$ 
   to obtain the desired 10 mRNAs per cell \cite{warren06_proc-natl-acad-sci-u}. Furthermore we set the protein degradation rate $\delta = 3.6\cdot 10^{-6} s^{-1}$ 
   as proteins are though to be one order of magnitude more stable than mRNAs \cite{Schwanhausser2011}. 
   The translation rate was set to $\beta = 0.0036\;\text{mRNA}^{-1} s^{-1}$ in accordance to \cite{Schwanhausser2011}.
   We observe the emergence of the same four attractors described in the main text and Fig.\;3 therein.
}
  \label{fig:10000Landscape}
\end{figure}
\newpage
\newpage
\begin{figure}[b]
  \centering
  \includegraphics[width=1\textwidth]{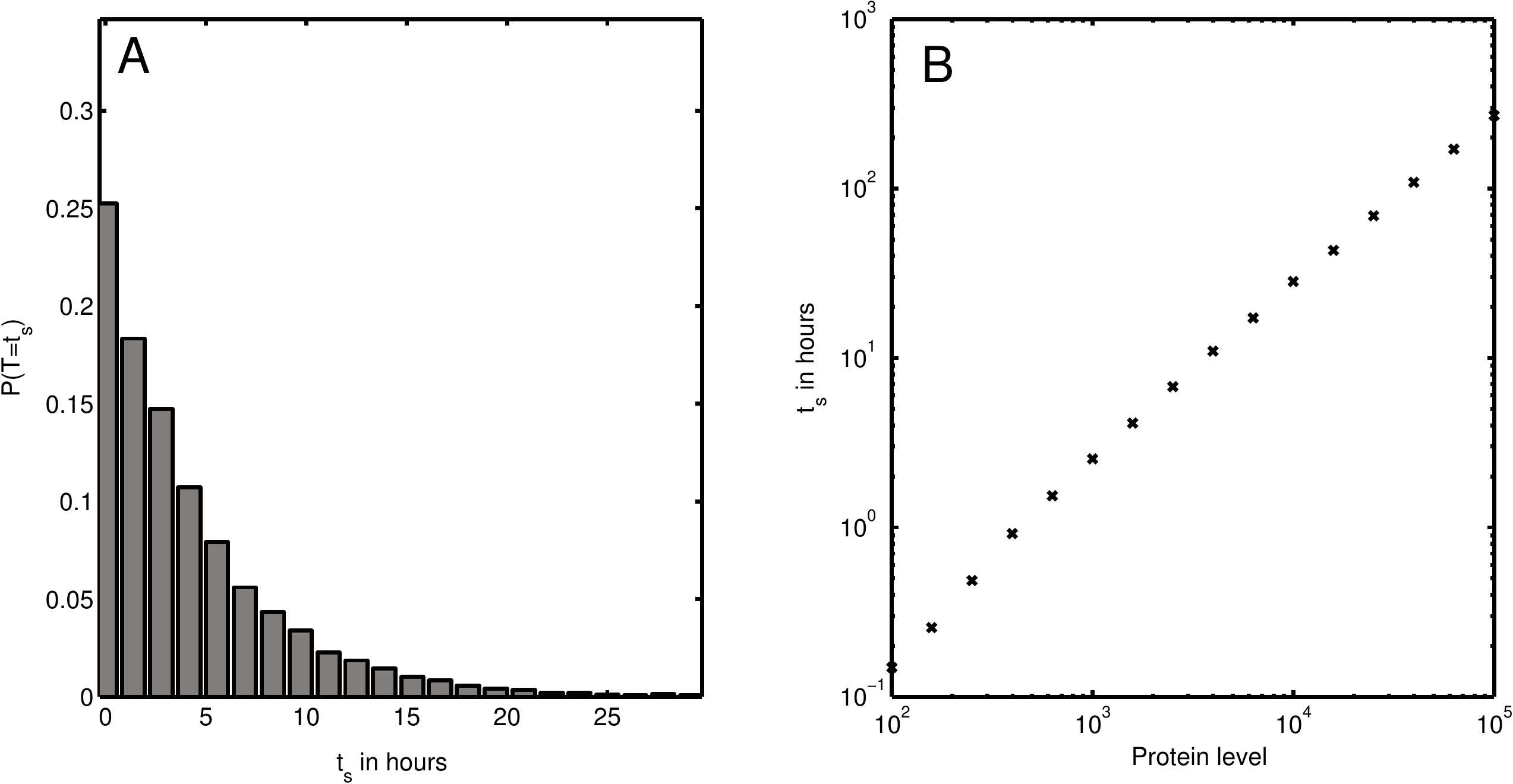}
  \caption{Residence time of the system in the intermediate attractors ($S^*_\playerOne$ or $S^*_\playerTwo$). 
(A) The histogram of the residence times in $S^*_\playerOne$ obtained from stochastic simulation suggests that the residence time in the intermediate attractors follows a geometric distribution.
(B) Similar to Fig.\;4 one observes a linear scaling of the mean residence time and the protein abundance. 
} 
\label{fig:primingtime}
\end{figure}